\documentclass[titlepage,oneside,12pt]{article}
\usepackage[tiny,rm,pagestyles]{titlesec}
\newpagestyle{trbstyle}{
	\sethead{Genser et al.}{}{\thepage}
    \setfoot{TRB 2022 Annual Meeting}{}{Original paper submittal}
}
\titleformat{\section}{\bfseries}{}{0pt}{\uppercase}
\titlespacing*{\section}{0pt}{12pt}{*0}
\titleformat{\subsection}{\bfseries}{}{0pt}{}
\titlespacing*{\subsection}{0pt}{12pt}{*0}
\titleformat{\subsubsection}{\itshape}{}{0pt}{}
\titlespacing*{\subsubsection}{0pt}{12pt}{*0}

\usepackage{ccaption}
\usepackage{amsmath}
\makeatletter
\renewcommand{\fnum@figure}{\textbf{FIGURE~\thefigure} }
\renewcommand{\fnum@table}{\textbf{TABLE~\thetable} }
\makeatother
\captiontitlefont{\raggedright \bfseries \boldmath}
\captiondelim{\;}

\usepackage[sort,numbers]{natbib}

\setcitestyle{round}

\usepackage[pagewise]{lineno}

\newread\somefile
\usepackage{xparse}
\newcounter{totalwordcounter}
\newcounter{wordcounter}
\newcounter{modifiedwordcounter}

\makeatletter
\NewDocumentCommand{\wordcount}{s}{%
  \immediate\write18{texcount -sum -1 \jobname.tex > count.txt}%
  \immediate\openin\somefile=count.txt%
  \read\somefile to \@@localdummy%
  \immediate\closein\somefile%
  \setcounter{wordcounter}{\@@localdummy}%
  \IfBooleanF{#1}{%
  \@@localdummy
  }%
}
\makeatother

\usepackage{totcount}
\regtotcounter{table}   
\regtotcounter{figure}  

\newcommand{\numberofwordsthejournalthinksforafigure}{250}
\newcommand{\numberofwordsthejournalthinksforatable}{250}
\newcommand{\numberofwordsthejournalthinksforthereferences}{500}
\newcommand{\numberofwordsyoudecidedtoremove}{-260}

\newcommand{\modifiedwordcount}{%
\wordcount*
\setcounter{modifiedwordcounter}{\value{wordcounter}}%
\addtocounter{modifiedwordcounter}{\numexpr\numberofwordsyoudecidedtoremove} %
\number\value{modifiedwordcounter}
\renewcommand{\modifiedwordcount}{\number\value{modifiedwordcounter}}
}

\newcommand{\totalwordcount}{%
\setcounter{totalwordcounter}{\value{modifiedwordcounter}}%
\addtocounter{totalwordcounter}{\numexpr\numberofwordsthejournalthinksforafigure*\totvalue{figure}}%
\addtocounter{totalwordcounter}{\numexpr\numberofwordsthejournalthinksforatable*\totvalue{table}} %
\addtocounter{totalwordcounter}{\numexpr\numberofwordsthejournalthinksforthereferences} %
\number\value{totalwordcounter}
\renewcommand{\totalwordcount}{\number\value{totalwordcounter}}
}

\usepackage{mathptmx}

\usepackage[T1]{fontenc}
\usepackage{textcomp}

\usepackage{hyperref} 
\usepackage{graphicx}
\usepackage{tikz}
\usepackage[utf8]{inputenc}
\usepackage{multirow}
\usepackage{pstricks}
\usepackage{xstring}
\usepackage{advdate}
\usepackage{amssymb}
\usepackage{amsthm}
\usepackage{verbatim}
\usepackage{enumitem}
\usepackage{booktabs}

\usepackage[labelformat=simple]{subcaption}

\usepackage{algorithm}
\usepackage{algpseudocode}
\usepackage{placeins}
\usepackage{tikz}

\definecolor{darkblue}{rgb}{0,0,0.5}

\hypersetup{colorlinks=true,linkcolor=black,citecolor=darkblue,urlcolor=darkblue} 

\newcommand{\newref}[2]{\hyperref[#2]{#1~\ref*{#2}}} 

\usepackage{varwidth}
\usepackage{array}
\newcolumntype{L}[1]{>{\raggedright\let\newline\\\arraybackslash\hspace{0pt}}m{#1}}
\newcolumntype{C}[1]{>{\centering\let\newline\\\arraybackslash\hspace{0pt}}m{#1}}
\newcolumntype{R}[1]{>{\raggedleft\let\newline\\\arraybackslash\hspace{0pt}}m{#1}}

\begin{document}

\thispagestyle{empty}


\begin{titlepage}
\begin{flushleft}

{\LARGE \bfseries An Experimental Urban Case Study with Various \\[0.2cm] Data Sources and a Model for Traffic Estimation}\\[0.8cm]

\textbf{Alexander Genser*}\\
Institute for Transport Planning and Systems, ETH Zurich\\
CH-8093, Switzerland\\
Tel: +41 44 632 75 19\\
Email: \href{mailto:gensera@ethz.ch}{gensera@ethz.ch}\\[0.5cm]

\textbf{Noel Hautle}\\
Institute for Transport Planning and Systems, ETH Zurich\\
CH-8093, Switzerland\\
Tel: +41 44 633 31 05\\
Email: \href{mailto:nhautle@student.ethz.ch}{nhautle@student.ethz.ch}\\[0.5cm]

\textbf{Michail Makridis}\\
Institute for Transport Planning and Systems, ETH Zurich\\
CH-8093, Switzerland\\
Tel: +41 44 633 31 05\\
Email: \href{mailto:michail.makridis@ivt.baug.ethz.ch}{michail.makridis@ivt.baug.ethz.ch}\\[0.5cm]

\textbf{Anastasios Kouvelas}\\
Institute for Transport Planning and Systems, ETH Zurich\\
CH-8093, Switzerland\\
Tel: +41 44 633 66 95\\
Email: \href{mailto:kouvelas@ethz.ch}{kouvelas@ethz.ch}\\[0.5cm]
 
* Corresponding author\\[0.5cm]



{\AdvanceDate[-1]\today}
\end{flushleft}
\end{titlepage}




\newpage

\thispagestyle{empty}
\section*{Abstract}
Accurate estimation of the traffic state over a network is essential since it is the starting point for designing and implementing any traffic management strategy. Hence, traffic operators and users of a transportation network can make reliable decisions such as influence/change route or mode choice. However, the problem of traffic state estimation from various sensors within an urban environment is very complex for several different reasons, such as availability of sensors, different noise levels, different output quantities, sensor accuracy, heterogeneous data fusion, and many more. To provide a better understanding of this problem, we organized an experimental campaign with video measurement in an area within the urban network of Zurich, in Switzerland. We focus on capturing the traffic state in terms of traffic flow and travel times by ensuring measurements from  established thermal cameras by the city's authorities, processed video data and the Google Distance Matrix. We assess the different data sources, and we propose a simple yet efficient Multiple Linear Regression (MLR) model to estimate travel times with fusion of various data sources. Comparative results with ground-truth data (derived from video measurements) show the efficiency and robustness of the proposed methodology. 

\vspace{2cm}
\noindent\textit{Keywords}: Urban traffic state, Travel time estimation; Traffic management; Traffic flow; License Plate Detection; Empirical measurements; Multiple linear regression. 
\newpage

\section{Introduction} 
\label{sec:intro}
 An accurate derivation of fundamental traffic state variables is key for sophisticated traffic management. Especially urban areas suffer from congestion due to a higher population density, traffic lights and elevated mobility demand. Besides other variables, traffic flow at specific locations in the network and travel times between an origin and a destination are of great importance for traffic operators. In particular, travel times allow for the derivation of the network's current Level of Service (LoS) and influence the network user's mode and route choice. Consequently, accurate sensor technology is needed to detect vehicles when traveling through a network with a low error rate that traffic variables satisfy accuracy requirements~\cite{ref:kouvelas_estimation}. However, urban areas are sometimes sparsely equipped with sensors, which negatively affects accuracy and increases the noise level of the results. In addition, a variety of sensor technologies used in urban environments are only suitable for deriving particular traffic variables and also differ in data resolution. For example traditional sensor technologies such as Loop Detectors (LD) are still one of the widely used measurement devices due to reliability and flexibility in design~\citep{ref:bachmann_fusing_bluetooth_LD}. Although, LDs have been shown as a promising data source for traffic flow, theoretical assumptions for a unique vehicle identification are needed and a re-identification is not possible. In contrast, new technologies such as video/thermal cameras and Bluetooth/WiFi sensors not only enable accurate derivation of traffic flow, but should also provide good results in measuring travel time, since unique vehicle identification based on MAC address recognition is possible.  
 
 Although numerous studies have evaluated and estimated travel times on freeways and in urban areas, not many studies have compared emerging sensor technologies to an empirical ground-truth measurement. Also, the question remains, how traditional sensor data can help with travel time estimation in terms of performance improvement. Therefore, this work focuses on providing traffic state representation in terms of traffic flow and travel time estimation within an urban network. We run an experimental campaign in Zurich, Switzerland to determine video data from the particular area and investigate the time series derivation of traffic flow and travel times. Consequently, we compare the following data sources: (a) thermal camera sensors data that are equipped with a WiFi interface (b) processed video data with an Automated License Plate Recognition (ALPR) algorithm and (c) Google Distance Matrix data from the particular area. For comparative results, the video data serves for the exact determination of traffic flow and travel times, i.e., a ground-truth data set. Besides the assessment of the different data sets, we propose a simple yet efficient Multiple Linear Regression (MLR) model to estimate travel times in a future environment with Connected and Automated Vehicles (CAV). We create a baseline scenario where 5\% of the data from moving sensors (e.g., from CAVs) is available and propose a model that fuses this data with traditional LD data.
 
 The paper includes the following contributions: (a) an experimental campaign with multiple sensor data (thermal video. inductive loop detectors, and Google Distance Matrix);
 (b) sensor-based analysis for traffic state estimation in terms of traffic flow and travel times;
 (c) accuracy assessment of the derived time series for all sensors; (d) travel time estimation for specific origin-destinations by fusing static LD data with (emulated) CAV information and 
 comparison to a  baseline scenario.
 
 The remainder of the paper is as follows: The next section highlights relevant previous works in the field of travel time assessment and estimation. A description of the empirical experiment and the collected data sets is given in the succeeding section. This includes a description of the area, the performed video measurement, and a description of the other sensor sources, i.e., thermal cameras, post-processed video data with ALPR, Google Distance Matrix data, and derivation of the ground-truth data set. Next, we introduce the data processing methodology and the utilized performance metrics. The travel time estimation model with used performance metrics is introduced in the following section. Afterwards, the results are presented and discussed and the paper closes with a concluding section and an outline for future work.

\section{Related works} 
\label{sec:relatedworks}
As a consequence of rising mobility demand, freeways and urban areas are continuously suffering from traffic congestion. This results in smaller network throughput, lower average speeds on network links, and consequently, higher travel times~\citep{ref:zheng_diss}. The traffic management domain offers several tools that influence route choice and mode choice (e.g., congestion pricing~\cite{ref:genser_pricing}), or traffic demand (e.g. user's departure time) to tackle rising congestion problems. Nevertheless, the implementation of sophisticated traffic management policies requires precise quantities as input, such as traffic flow and travel times. Besides, travel time estimation in urban networks is particularly challenging because of the dynamic demand, low speeds and signaling. Efforts towards accurately predicting time-to-green~\cite{ref:genser_T2G} can certainly help but until high penetration levels of AVs~\cite{ref:chavoshi_LI}, traffic signaling remains a challenge for accurate travel time estimation.

In particular, travel times can be measured directly by identifying a vehicle at two specific points in space with a corresponding timestamp. This is achieved with camera data and the application of ALPR algorithms that allows matching license plates or probe vehicle data. Also, novel sensors with Bluetooth or WiFi interfaces can detect a unique MAC address of devices such as, e.g., a mobile phone~\citep{ref:yildirimoglu_part_B_experienced_TT}. An evaluation of Bluetooth sensors as a potential ground-truth data source was performed in~\cite{ref:TT_data_collection}. The study compares Bluetooth sensor data collected on a freeway to vehicle probe data, and the penetration rate of the Bluetooth system is approximated with LD data. Results underline that Bluetooth sensors are a promising data source to measure accurate travel times.
Nevertheless, the quality of Bluetooth sensor data heavily depends on the penetration in the system. Hence, the statement that the new sensors can be utilized as ground-truth data is questionable with a low sample rate. As~\cite{ref:sharifi_bluetooth_eval} shows, the penetration rate of Bluetooth systems can be low and hence only detect a fraction of vehicles of an observed network. The study reports that the aggregated penetration rate of several observed segments in Maryland, USA, fluctuates around 4\%. For the derivation of the ground-truth data set, LDs and microwave sensors are utilized; i.e., no actual empirical measurement of the traffic flow or travel time was performed. The results are supported by another study in Turkey, where a penetration rate of 5\% was found~\cite{ref:erkan}. Another field test was conducted by~\cite{ref:barceloe} on a freeway in Barcelona, Spain. The work utilizes Bluetooth sensors to determine travel times and then designs a Kalman filter that allows the estimation of origin-destination pairs on freeways. The authors extend their work in~\cite{ref:barceloe2} to urban networks with route choice. 

Alongside the determination of travel times for vehicles on freeways and urban areas, the quantities are also from great interest for public transportation, non-motorized transport modes, i.e., cyclists and pedestrians.~\cite{ref:elliott_TT_modeling} develops a framework to estimate the travel times of buses in Auckland, New Zealand. The framework allows for calculating a network's traffic state, i.e., congested or not congested with positioning data. Two models are combined, where the first model estimates travel times followed by the traffic state computed by the second model.

Contrary, travel times can also be derived indirectly with traditional sensors data. Loop Detectors allow the direct determination of traffic flow and an approximation of speed, which can then be used to calculate travel times. A prediction of travel times based on LD data in California is proposed by~\cite{ref:yildirimoglu_part_B_experienced_TT}. The processed data serves as input to a prediction framework, including a bottleneck identification algorithm, traffic regime clustering, a stochastic congestion map for clustered data, and an algorithm for a congestion search algorithm. A more practical approach of travel time estimation was followed by~\cite{ref:shen_practical_TT_estimation} with LD data. The work addresses data gaps (spatially and temporally) by applying an Exponential Moving Average (EMA) to improve the estimates.  Additionally,~\cite{ref:shen_practical_TT_estimation} transforms the calculated Time-Mean-Speed (TMS) to Space-mean-speed (SMS) by fitting a linear regression; the transformation improves the estimates further. However, the authors do not provide details about their models and only utilize simulated data from Aimsun (i.e., no ground-truth data) to validate their approach. The study in~\cite{ref:yeon_part_B_TT_estimation} utilizes LD data to calculate speed, traffic volume, and the LD's occupancy. The acquired data serves to validate a developed freeway travel time estimation model with Discrete-Time Markov Chains. The model shows promising results with travel time deviations lower than 3\% from the LD data. Nevertheless, a comparison to an actual empirical ground-truth measurement is not provided. 


\section{Description of experimental campaign and data sources} 
\label{sec:perimeter}
As this study focuses on traffic state estimation via traffic flow and travel time derivation in an urban environment, a small area is selected that reflects the quantities variance but does not allow for complex traffic movements, e.g., a high number of possible routes. The particular area is located in Zurich, Switzerland in the northern city part. The experimental campaign to compile video data is described in the following; the various data sources utilized here for quality assessment are also introduced in the next section. 

\subsection{Experimental campaign with video cameras}
\label{sec:exper_campaign}
\begin{figure}[!t]
    \centering
    \includegraphics[width=0.8\textwidth]{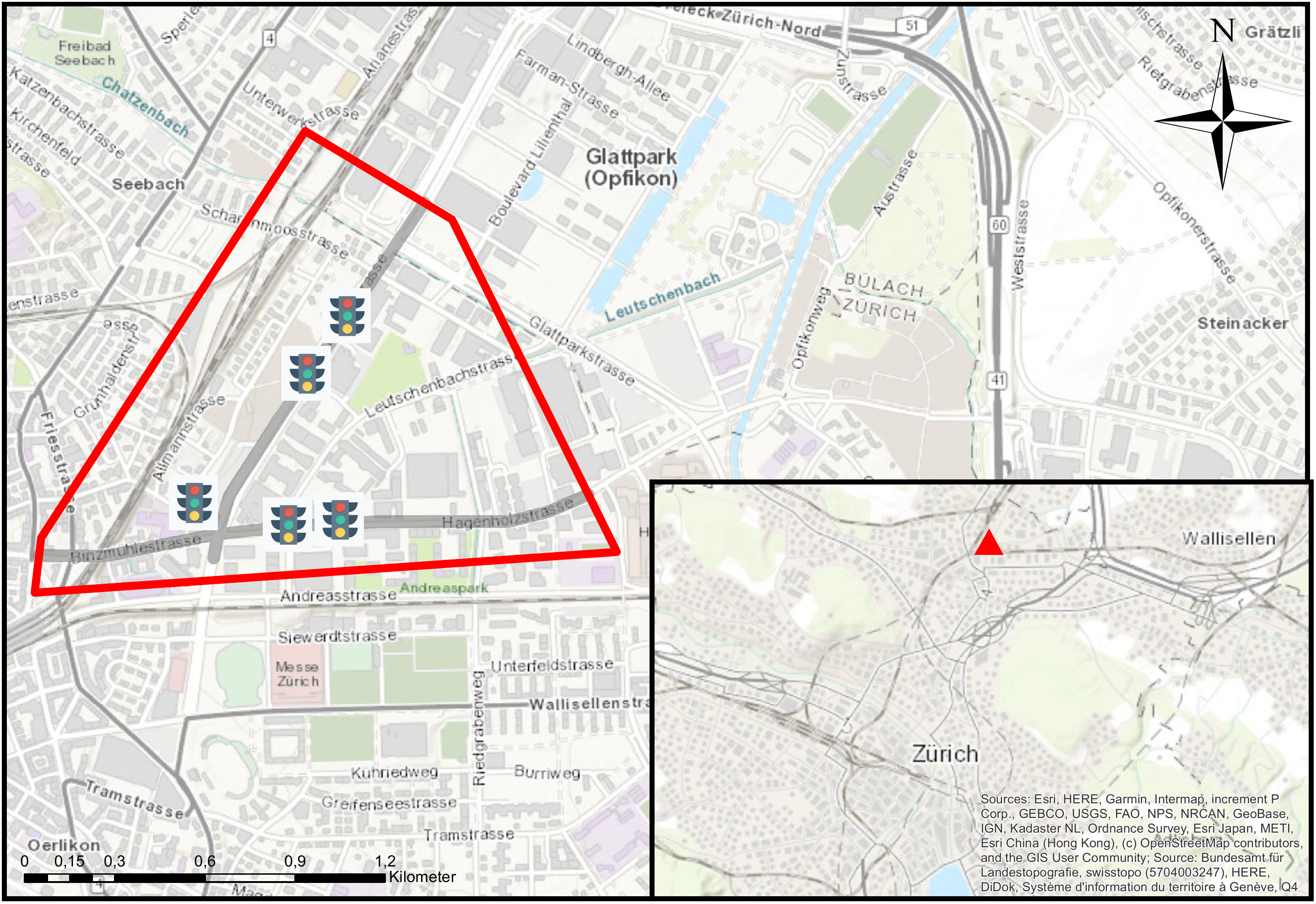}
    \caption{Test area (in red), Zurich Switzerland. The highlighted network in gray indicates the road segments where data is collected and processed. The traffic light symbols indicate the five implemented traffic control systems.}
    \label{fig:perimeter}
\end{figure}

Figure~\ref{fig:perimeter} depicts the area under investigation in red. Additionally, the road sections that are investigated with different data sources are highlighted in gray. The approach starting in the west named Binzm{\"u}hlstrasse is denoted as West bound (WB); the east approach called Hagenholzstrasse as East Bound (EB); the approach from the north named Thurgauerstrasse as North Bound (NB). The transportation network in the red area primarily serves individual motorized transport and is regulated with five traffic control systems. A bus line is operating on the west-east axis (WB to EB) and vice versa (EB to WB). Bicycle traffic is managed with cycle paths implemented on the road or with separated cycling infrastructure.

\begin{figure}[!t]
    \centering
      \begin{subfigure}[b]{0.49\textwidth}
      \centering
       \includegraphics[width=1.05\textwidth]{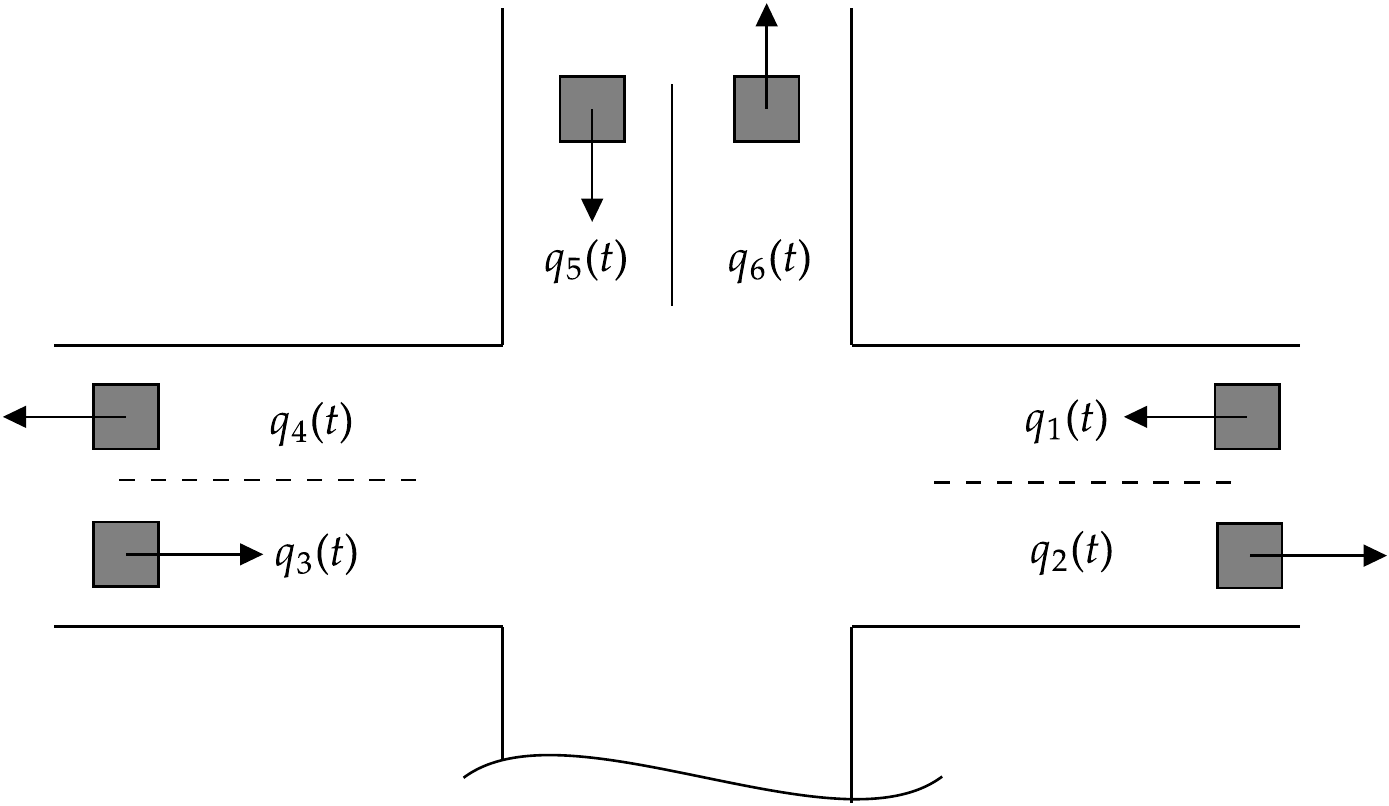}
        \caption{}
        \label{fig:flow_detection_spots}
    \end{subfigure}
    \begin{subfigure}[b]{0.49\textwidth}
    \centering
       \includegraphics[width=0.95\textwidth]{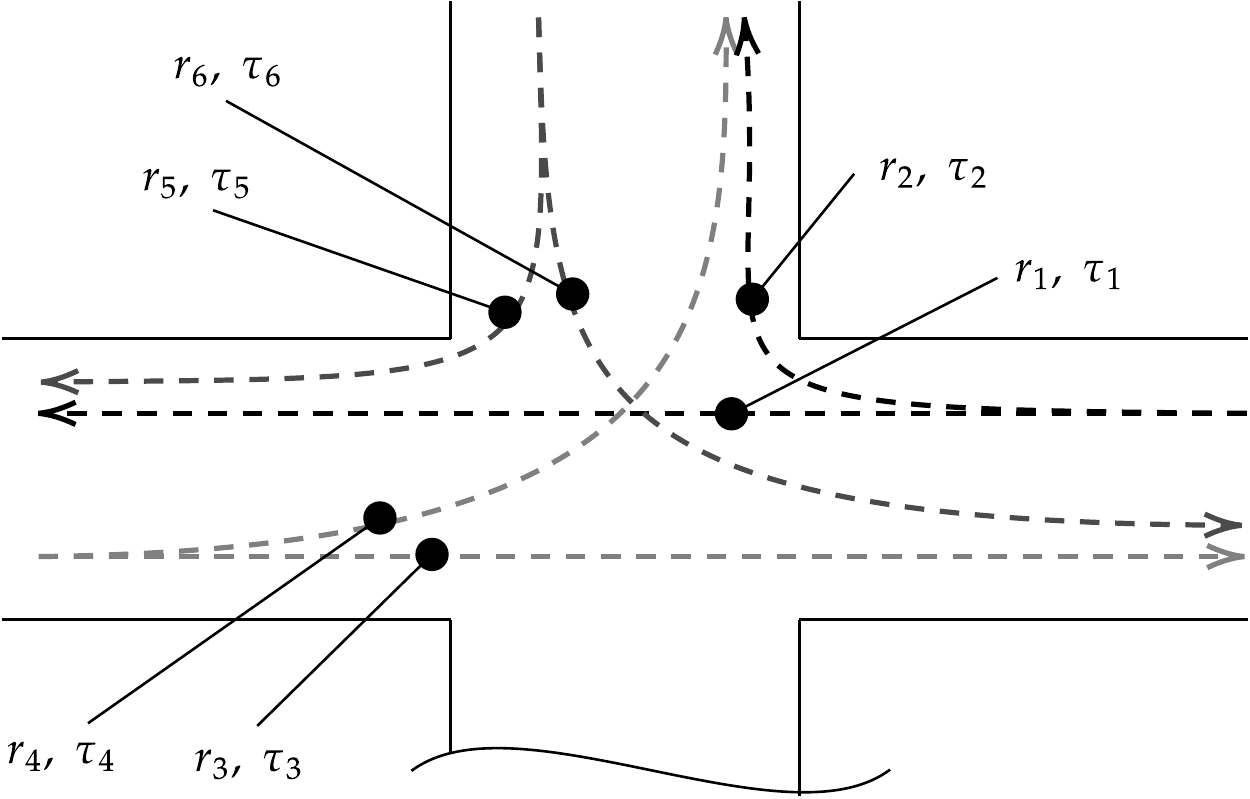}
        \caption{}
        \label{fig:travel_time_routes}
    \end{subfigure}
    \caption{Defined quantities for derivation of traffic flow and travel time: (a) measurement spots to derive traffic flow $q_s(t)$ for all $s=\{1,2,3,4,5,6\}$, (b) pre-defined routes $r=\{1,2,3,4,5,6\}$ to derive all travel times $\tau_r(t)$. }
    \label{fig:spots_routes}
\end{figure}

Although Zurich shows good coverage of various sensors over the whole city, the area is selected for the following reasons: (a) the deployed sensors are spatially close to each other. Thus, the complexity is adequately low, e.g., the number of allowed traffic movements and the number of intersections. (b) the installed sensors allow the observation of three out of four intersection approaches. (c) the area shows good coverage with loop detector and signals control devices utilized for the feature extraction for travel time estimation. Consequently, the fundamental quantities traffic flow $q_s(t)$ of pre-defined points $s$ in the area space and travel time $\tau_r(t)$ for a pre-defined route $r$ are defined. Figure~\ref{fig:spots_routes} depicts all spots $s$ and routes $r$. Note that the approach from the south is not defined, and consequently, no elements of $s$ and $r$ are available. Since no sensors are deployed on this road section, an in-detail investigation is not possible. Consequently, six measurement spots to derive a set of traffic flow $\mathcal{F}=\{q_1(t), q_2(t), q_3(t), q_4(t), q_5(t), q_6(t)\}$ and a set of six routes $\mathcal{R}=\{r_1, r_2, r_3, r_4, r_5, r_6\}$ for derivation of the set $\mathcal{T} = \{\tau_1(t), \tau_2(t), \tau_3(t), \tau_4(t), \tau_5(t), \tau_6(t)\}$, denoting the travel times throughout this work. 

\begin{figure}[!t]
    \centering
      \begin{subfigure}[b]{0.49\textwidth}
      \centering
       \includegraphics[width=0.9\textwidth]{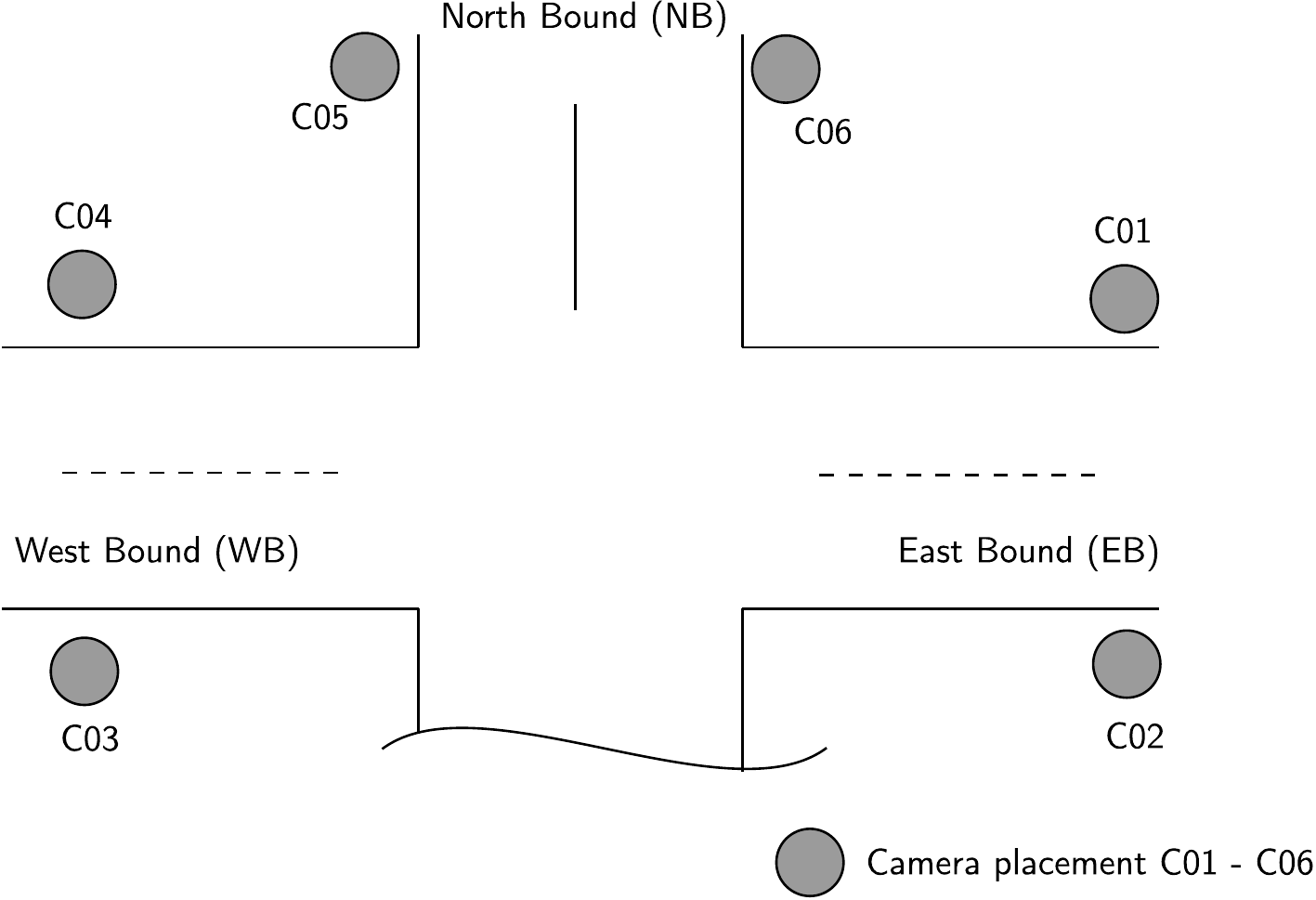}
        \caption{}
        \label{fig:cam_placement}
    \end{subfigure}
    \begin{subfigure}[b]{0.49\textwidth}
    \centering
       \includegraphics[width=0.9\textwidth]{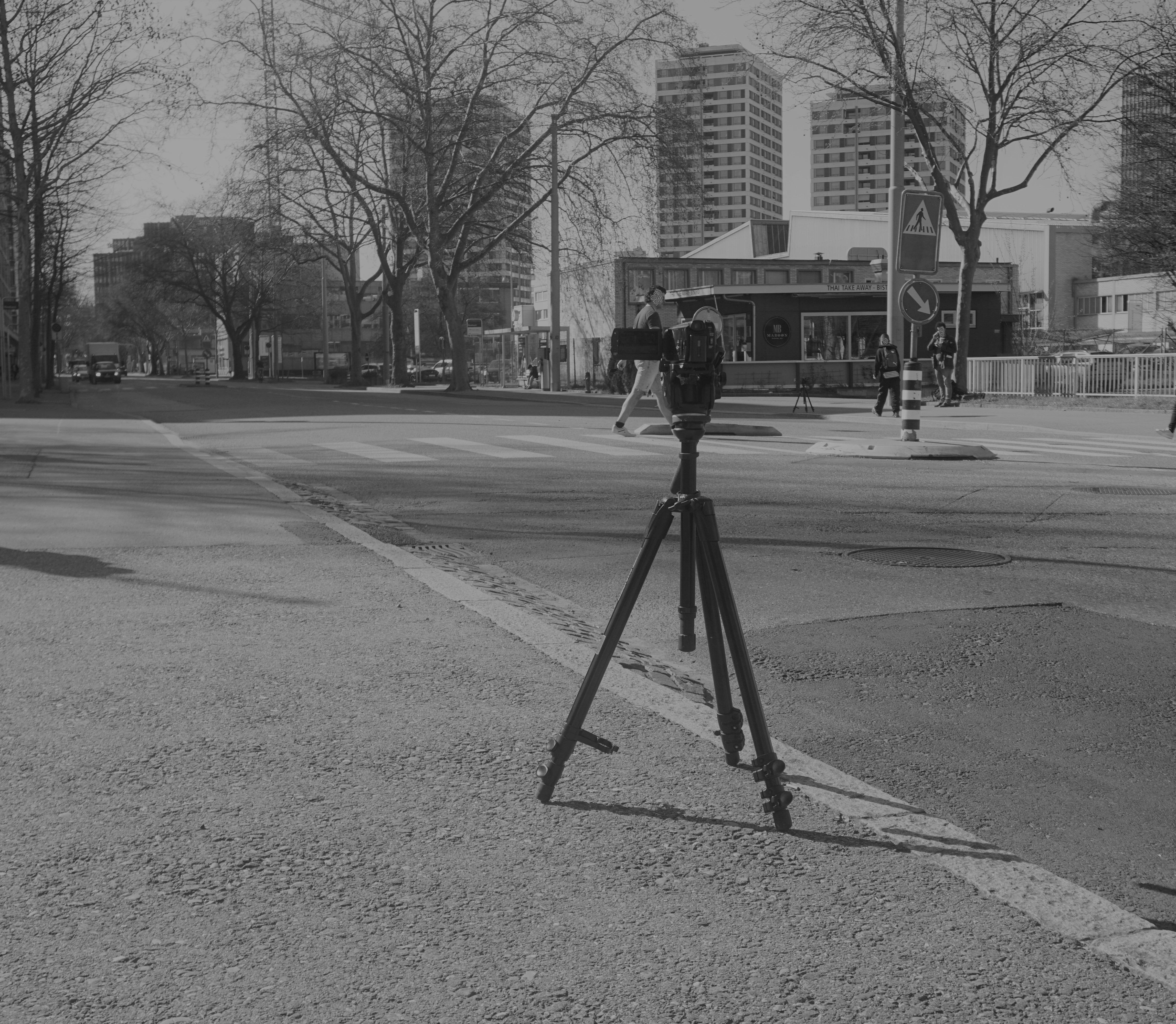}
        \caption{}
        \label{fig:pics_cam}
    \end{subfigure}
    \caption{Set-up for empirical measurements: (a) the placement of six cameras to capture traffic traversing (C01 -- C06), (b) used camera set-up with HD camera and tripod.}
    \label{fig:cameras}
\end{figure}

For carrying out the empirical experiment, a prior analysis of traffic data from~\cite{ref:TomTomMethodology} was performed that proved the existence of a morning and evening peak hour. However, the evening peak hour showed more severe congestion. It was decided to perform the measurement in the period 4:00 pm -- 6:00 pm, i.e., traffic movements are filmed for two hours on March 31, 2021. To determine the video data for all traffic flows $q_s(t)$ and travel times $\tau_r(t)$ measurements with video cameras were performed. Six HD cameras with tripods were used and placed according to a prior-analysis of adequate measurement spots to cover all three traffic axis in the area. The cameras are observed by one person each to ensure a correct and precise measurement procedure. The spots C01 -- C06 are depicted an Figure~\ref{fig:cam_placement} and the camera set-up in Figure~\ref{fig:pics_cam}. For the positioning of the cameras, the following conditions were ensured: No influence of traffic flow or traffic behavior; no objects crossing the camera image such as, e.g., pedestrians; a camera angle that allows minimization of disturbances due to, e.g., light reflections. A camera positioning that ensures that the conditions hold allows eliminating error sources in the automated video processing carried out later in this work. 

\subsection{Data sources for sensor assessment}
\label{sec:data_sources}
To develop a sensor-based analysis for traffic state representation in terms of traffic flow and travel time and provide an accuracy assessment, the following data sources are introduced in this section: already established thermal sensors in the area, processed video data with the application of an ALPR algorithm, a compiled data set from the Google Distance Matrix, and the manual inspection of the video data to derive the true traffic flow and travel times.

Due to several disadvantages of visual cameras (e.g., privacy concerns, lightning disturbances, visibility of objects, no observability of objects during the night), recently, thermal cameras became more popular to observe traffic. 
Additionally, camera sensors (visual or thermal) also have been combined with Bluetooth or WiFi sensors that allow capturing the signals of mobile devices. When a mobile device is detected, a unique MAC address can be identified that enables vehicle identification and re-identification~\citep{ref:Ding2019EvaluationTest}. By matching the unique MACs occurring at two or more sensor locations in a network, travel times and the average speed can be determined \citep{ref:Kwong2009ArterialSensors}. 

\begin{figure}[!b]
    \centering
      \begin{subfigure}[b]{0.49\textwidth}
      \centering
       \includegraphics[width=0.9\textwidth]{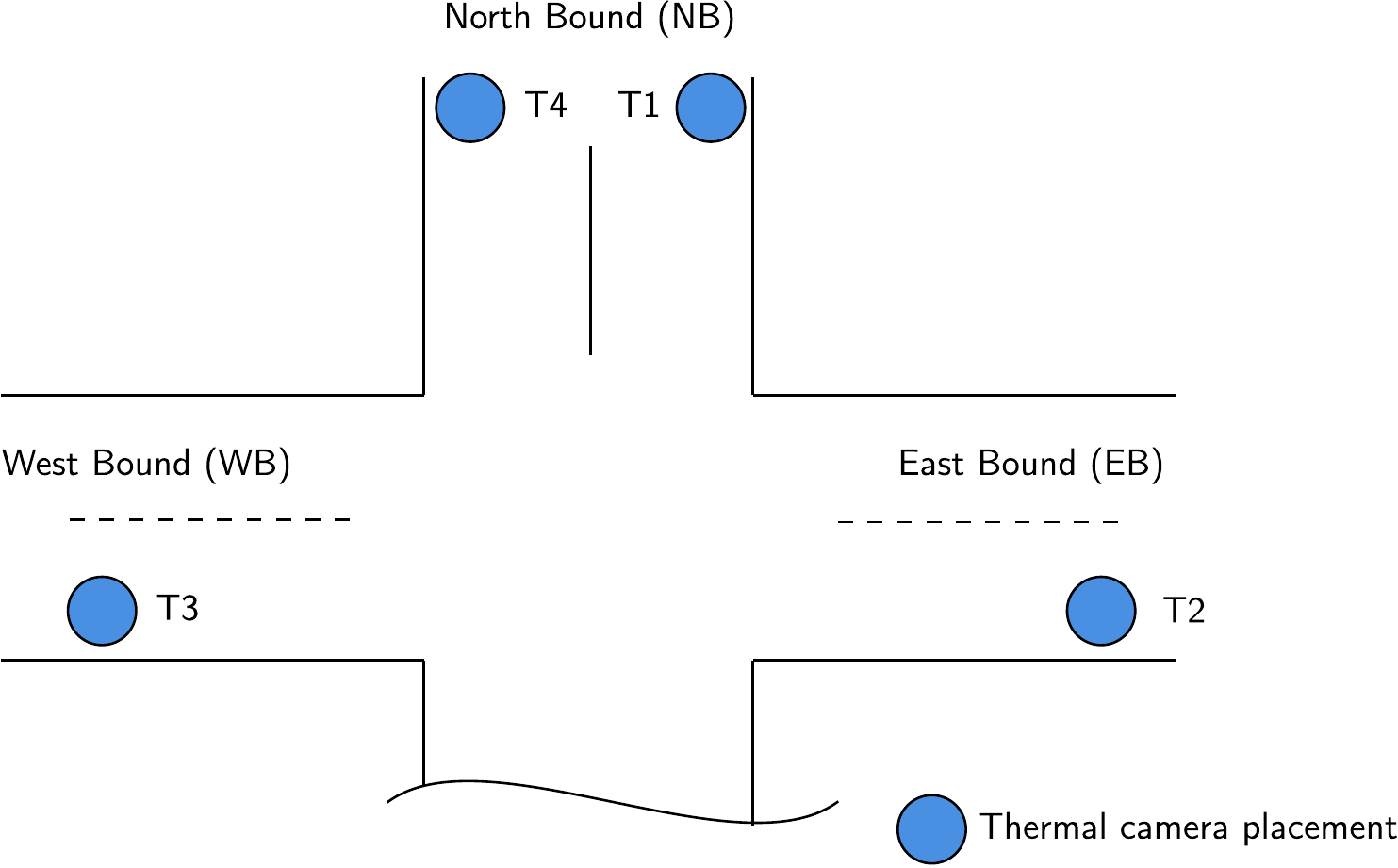}
        \caption{}
        \label{fig:th_cam_placement}
    \end{subfigure}
    \begin{subfigure}[b]{0.49\textwidth}
    \centering
       \includegraphics[width=0.6\textwidth]{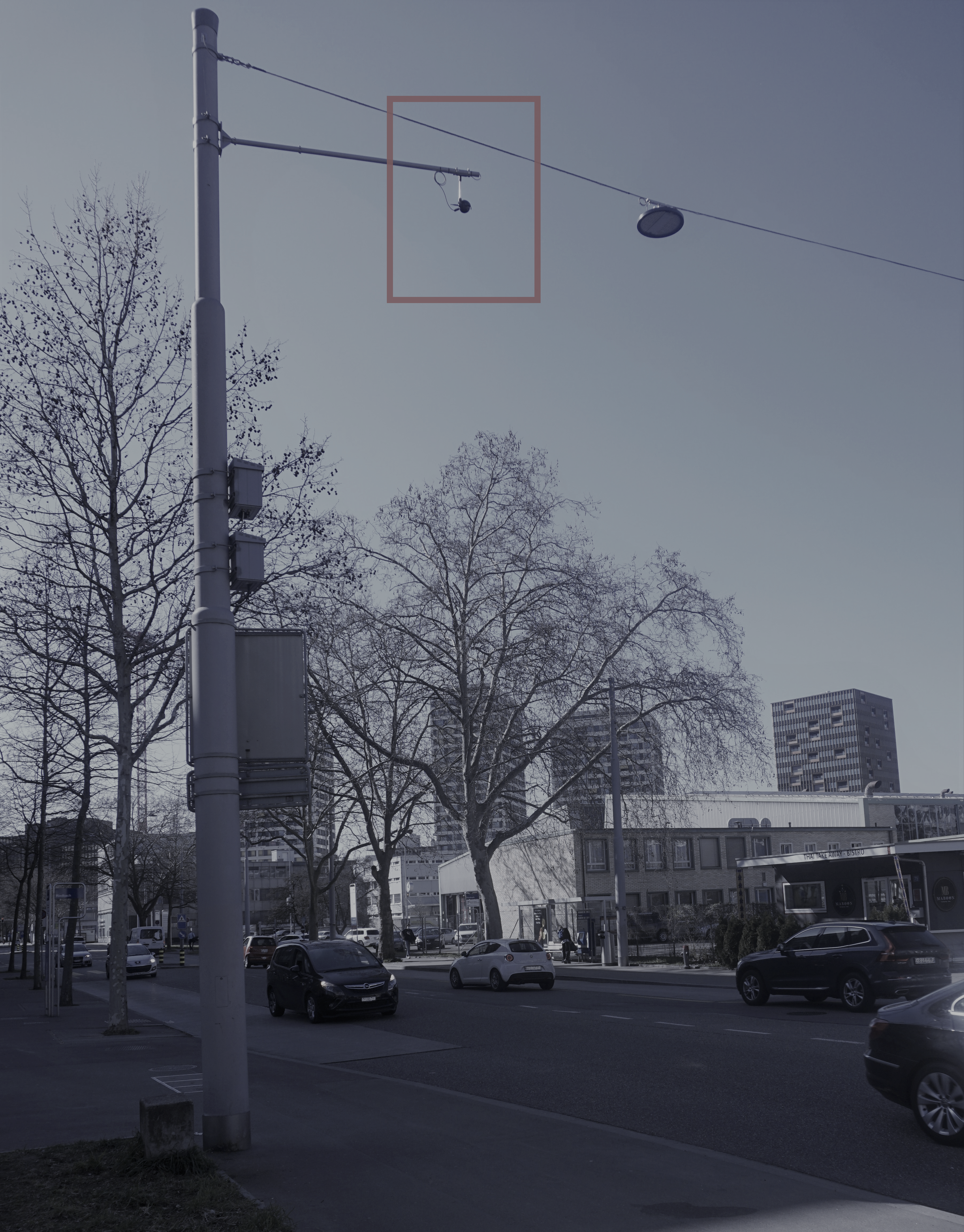}
        \caption{}
        \label{fig:th_pics_cam}
    \end{subfigure}
    \caption{Thermal camera set-up: (a) the mounted four thermal cameras to capture traffic traversing (T1 -- T4), (b) example of a overhead mounted thermal camera.}
    \label{fig:th_cameras}
\end{figure}

The area under investigation is equipped with four overhead thermal cameras as depicted in Figure~\ref{fig:th_cameras}, T1 -- T4. Note that the T2 and T3 capture both directions of traffic. The camera technology is capable of detecting vehicles with user-defined virtual detection zones. Thus, the determination of traffic flow for the set $\mathcal{F}$ is possible. Additionally, the camera detects WiFi signals which allow the derivation of travel times $\mathcal{T}$ in the area. For the export of the data, a commercial software system deployed by the camera manufacturer is used. This should allow practical insights into the data quality practitioners can expect when using such techniques for traffic management. 

To show the capabilities of the empirical video data, we apply an ALPR algorithm to post-process the recorded frames. ALPR is the automated process of recognizing an LP and matching the correct letters and numerical characters. Different research areas have proposed a variety of open source and commercial solutions for ALPR~\citep{ref:bakhtan_lr_alpr}. However, such algorithms are often tested under idealized conditions (e.g.,no reflections due to sunlight).
We apply a novel algorithm from~\cite{ref:silva2018a} to our video data set to determine traffic flow and travel times. The algorithm applies YOLO to detect objects as a first step. As we ensure that only vehicles are detected, we can use the count as a proxy to determine traffic flow. Travel times require a few more processing steps as we are the first to apply this algorithm in Switzerland (i.e.\ filtering different LP formats). When a LP is identified as correct by the filtering procedure, the timestamp of entering or exiting a route $r$ is stored in the data set. Again, for the further data processing steps to determine $q_s(t) \forall s$ and $\tau_r(t) \forall r$, please see the methodology described in the following section. 

Google allows requesting travel distance and travel time data via the Google Distance Matrix API. The API returns data based on HTTPS request information, including a start and end location (specified as GPS points). For full documentation and parameterization of the API, the interested reader is referred to~\cite{ref:google}. In this work we first specify the set of routes $\mathcal{R}$ with a start and endpoint as latitude and longitude coordinates. In addition, the heading angle of traffic is parameterized (i.e., for $r_1$ the angle, is equal to 90 degrees, indicating traffic moving east). Finally, we collect travel times of all routes for the transportation mode 'driving' and the traffic model 'best\_guess'. 
As the Google Distance API only allows to request data for the current timestamp or travel time predictions, historical data can not be accessed. Therefore, we could not obtain the data set for the time frame, we performed the empirical measurements with the video cameras. Hence, we developed an algorithm that tracks travel times for all routes every Wednesday for the identical time frame the empirical measurement was performed. Consequently, we build up a data set with Google Distance Matrix travel times of three Wednesdays. The developed algorithm sends HTTPS-requests to the Google API, post-processes the response and adds a new entry to the data set. Assuming that traffic demand in the area is similar every Wednesday from 4 pm -- 6 pm, we compare the average of the collected travel times to the other data sources. Note that this data set only allows the determination of travel times, and that is not further processed with the methodology presented in the next section.  

Finally, the video data was processed manually to compile the ground-truth data set. The two hours of video material are inspected by hand, and (a) vehicles were counted per minute, and (b) the time stamp a vehicle passes a spot $s$ of every intersection approach was determined. Note that due to privacy regulations, the data set required the deletion of all LPs after the data collection; i.e., we assigned a unique ID to every vehicle in the system. For the further data processing steps to determine $q_s(t), \ \forall s$ and $\tau_r(t), \ \forall r$, please see the methodological part that follows.

\section{Methodology for representing urban traffic states}
\label{sec:methodology}
This section presents the methodology to infer traffic flow and travel times from the available data sets. In addition, we also introduce performance metrics that are used to evaluate the derived time series. First, we derive traffic flow in (veh/h) from post-processed video data with the ALPR algorithm and manually from the video data set. The count data was aggregated to a resolution of one minute for all locations. However, the count data is then scaled that the flow time series shows values in vehicles per hour. Thus, we can define $q_s(t)$ as follows:
\begin{equation}
    q_s(t) = \frac{n}{T},
\end{equation}
where $n$ represents the vehicle count, and $T$ represents the measurement interval, which is constant for all sensor sources. Since the derived flow can show outliers or anomalies, a moving average of the data is derived as follows: 
\begin{equation}
    \Bar{q}_s(t) = \sum_{i = -k}^{0} q_s(t+i), 
\end{equation}
where $\Bar{q}_s(t)$ is the averaged traffic flow time series for $s$ and $k$ the window size specifying the data taken into account for the moving average of every sample of $\Bar{q}_s(t)$. Note that the defined window only considers historical data; this definition helps for real-time applications, where no future values are available. If the number of available samples are less than $k$ no averaged sample is computed. 

To derive the travel times, we again take the results from the ALPR and the ground-truth data set. For both, the travel time is derived by applying Equation~\ref{eq:tt}. The time series of travel time $\tau_r(t)$ for a route $r$ is defined as follows:
\begin{equation}
\label{eq:tt}
    \tau_r(t) =  \frac{1}{N}\sum_{v = 1}^{N} t_{v_\mathrm{out},r} - t_{v_\mathrm{in},r},
\end{equation}
where $t_{v_\mathrm{in},r}$ denotes the timestamp $t$ of a vehicle $v$ entering the system and following a route $r$. $t_{v_\mathrm{out},r}$ denotes the timestamp $v$ finished route $r$, i.e, exits the system. To create the travel time series, we average the number of vehilces $v$ denote as $N$ that travel $r$ at $t$. Again, to remove outliers, the travel time series are post-processed. As $\tau_r(t)$ is dependent on the $\Bar{q}_s(t)$, we chose to compute the weighted moving average, where $\Bar{q}_s(t)$ represents the weight, i.e., 

\begin{equation}
     \Bar{\tau}_r(t) =  \sum_{i = -k}^{0} \Bar{q}_s(i)\cdot \tau_r(t+i). 
\end{equation}
$\Bar{\tau}_r(t)$ represents the weighted moving average of $\tau_r$. Note that the data sets from the thermal cameras and the Google Distance Matrix contain already post-processed travel times. 

For the sensor-based assessment, we utilize two common performance metrics to compare the introduced data sources (for traffic flow and travel times, respectively). First, the Pearson correlation coefficient (PCC) is introduced as follows: 

\begin{equation}
    \rho(a,b) = \frac{E(a,b)}{\sigma_a \sigma_b},
\end{equation}
where $\rho(a,b)$ denotes the PCC between variables $a$ and $b$; $E(a,b)$ the cross-correlation between $a$ and $b$; $\sigma_a$ and $\sigma_b$ the variance of the variables, respectively. Thereby, a PCC of 1 indicates perfect postive correlation, -1 a perfect negative correlation, whereas a correlation coefficient of 0 indicates no correlation. The closer PCC to 1, the higher the correlation of the two variables~\cite{ref:PCC}.
Secondly, we use the Mean Absolute Percentage Error (MAPE), defined as
\begin{equation}
  \label{eq:MAPE}
    \mathrm{MAPE} = \frac{1}{n} \sum_{i=1}^{n} \left| \frac{y_i - \hat{y_i}}{y_i} \right|, 
\end{equation}
where $y_i$ are the actual observations of the ground truth, and $\hat{y}_i$ is the derived traffic quantities of a data source under evaluation~\citep{ref:MAPE}.

\section{Travel time estimation methodology}
\label{sec:TT_methodology}
For evaluating the hypotheses, if it is possible to estimate the travel times of an origin-destination pair in the area with multiple data sources, we design an MLR model for prediction. The goal is to design a model that can be utilized for prediction on all available the routes. The data taken into account is (a) a random 5\% data sample available from moving sensors in the area and (b) processed LD and traffic signal data from the intersections' signal controls within the area. Note that the 5\% random sample can be interpreted as probe-vehicle data or -- considering future transportation systems -- CAVs that transmit, e.g., location or speed, to a centralized processing unit. For the LD and signal data, 48 different loop detectors and traffic signals are considered for the regression model. For evaluation, if the LD and signal data improve the estimation model, a baseline model only with the 5\% ground-truth sample as a predictor is computed for every route. Secondly, the final prediction model is determined with all the designed features by performing an iterative process of forward selection for the predictor variables. The procedure of defining the MLR, the utilized features, and finding the final model are presented in the next section. 

\subsection{Model specification}
\label{sec:model_spec}
A mathematical formulation of the an MLR model can be introduced as follows:
\begin{equation}
    \begin{split}
         \hat{y}_{r,s} = \beta_{r,0} + \beta_{r,1} x_{s,1} + \beta_{r,2} x_{s,2} + \ldots \\
          + \beta_{r,p} x_{s,p} + E_{r,s}, \ \forall s \in \{1, \ldots, T\},
    \end{split}
    \label{eq:LR}
\end{equation}
where $\hat{y}_{r,s}$ is the response variable, i.e., the travel time estimate for a route $r$ and sample $s$. The intercept of the MLR equation is denoted as $\beta_{r,0}$; $\beta_{r,1} $ to $\beta_{r,p}$ denote all the regression coefficients for $r$ and a number of $p$ predictors. The predictors are denoted as $x_{s,1}$ to $x_{s,p}$. The error term is denoted by $E_{r,s}$ and follows a Gaussian distribution and $T$ denotes the prediction horizon.  The definition of the model is based on works such as~\cite{ref:LR1}. The solution for $\hat{y}_{r,s}$ is found by applying the Ordinary Least Square method.
First, the model only with the 5\% data sample as a predictor is designed for all routes $r$ as a baseline to compare the travel time estimates of the final selected model; i.e., the model definition in~(\ref{eq:LR}) only has one predictor. The baseline model is extended by features extracted from the LD and signal data for the final model. Besides, a non-linear variable transformation by utilizing the $\log(\cdot)$ function is used to change the regression relation and find the model with the best performance. The features and the corresponding transformation are as follows:  

\begin{itemize}
    \item Average headway (s) when a traffic light is green; Variable transformation = $\log(\cdot)$
    \item Progressed flow at an intersection (veh/h); Variable transformation = none
    \item Average occupancy of LDs (\%); Variable transformation = none
    \item Average red/green phase count (-), Variable transformation = $\log(\cdot)$
\end{itemize}
In addition, the response variable, i.e., the travel time for a route is also transformed with $\log(\cdot)$. 
To perform the training and validation of the model, the data is split into a training and test data set. 70\% of observations were used for training the model, and the last 30\% for testing. For the different routes, between 76 and 82 observations were available for the training, and 33 to 35 observations to validate the prediction results (test data set).

\subsection{Performance metrics}
\label{sec:performance_MLR}
Two performance metrics are utilized to evaluate the performance of the baseline and final determined MLR model.  
The coefficient of determination ($R^2$) expresses the percentage of the response variables' total variation that is explainable by the predictor variables. 
However, higher values of the $R^2$ measure can occur when more predictor variables are added to the model, leading to a distorted comparison. To account for this problem, the adjusted $R^2$ ($\mathrm{adj}R^2$) is introduced. $\mathrm{adj}R^2$ uses a penalty term that is dependent on the number of predictor variables ($p$). 
Additionally, the $\mathrm{MAPE}$ is utilized again as a measure to evaluate the travel time estimates of the model (i.e., the prediction compared to the test data set). 

\section{Results}
\label{sec:results}
In this section, the sensor-based assessment of traffic flows and travel times are shown in the first subsection. The second subsection presents the estimation results of the final MLR model and compares performances with the defined performance metrics. 
 
 \subsection{Traffic state representation and sensor-based assessment}
 \label{sec:results_traffic_state_understanding}
\begin{figure}[!b]
    \centering
      \begin{subfigure}[b]{0.49\textwidth}
      \centering
       \includegraphics[width=1\textwidth]{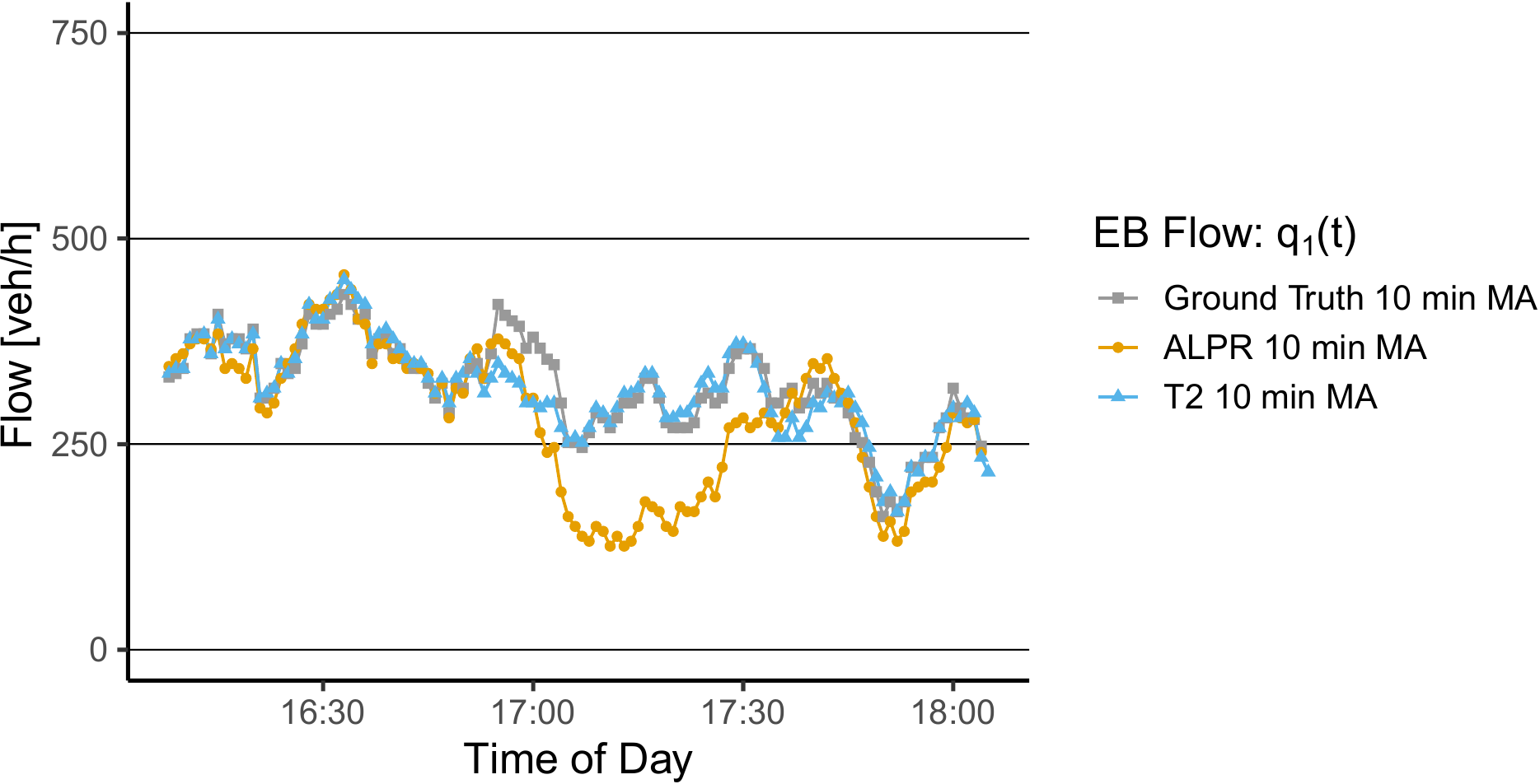}
        \caption{}
        \label{fig:q1}
    \end{subfigure}
    \begin{subfigure}[b]{0.49\textwidth}
    \centering
       \includegraphics[width=1\textwidth]{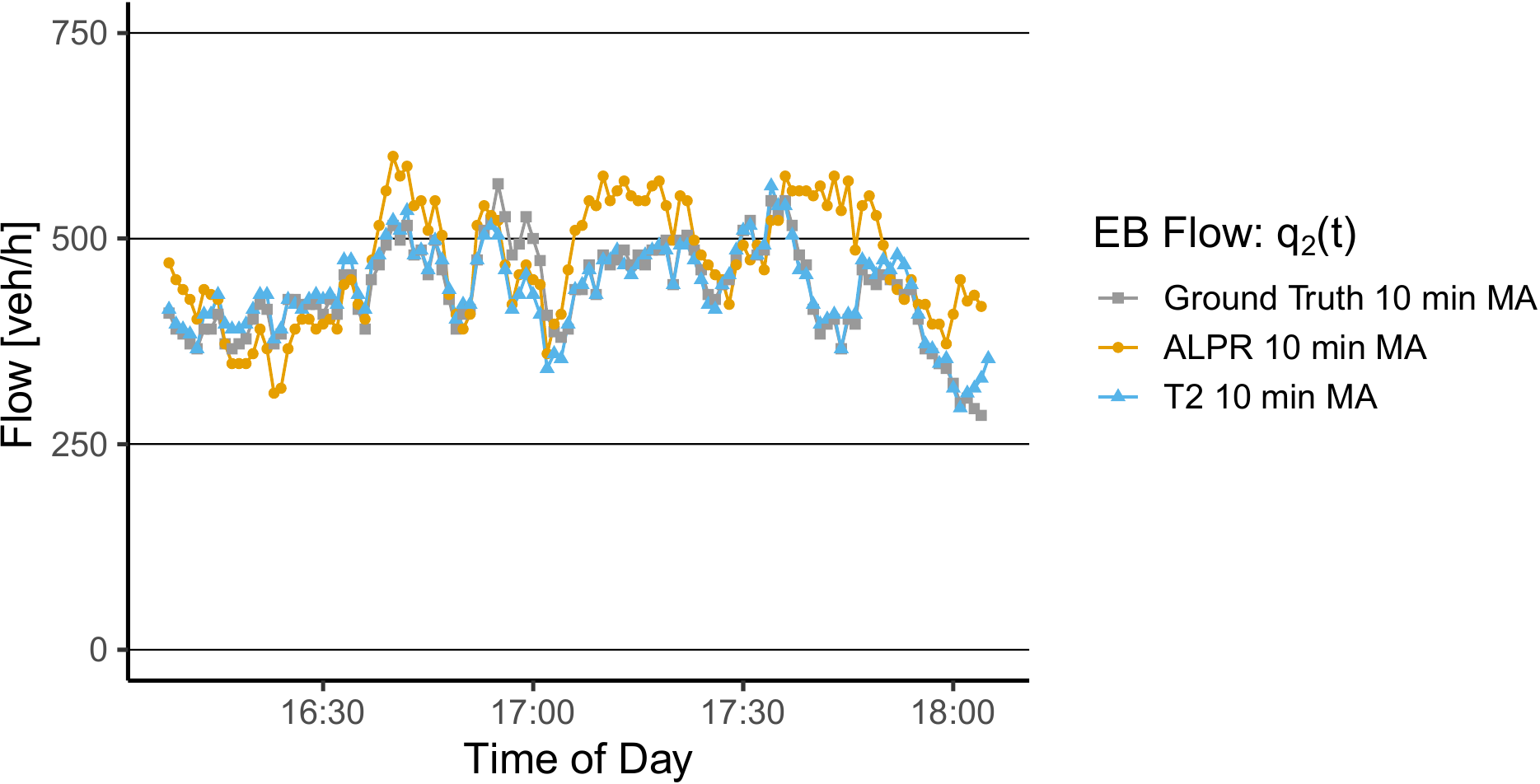}
        \caption{}
        \label{fig:q2}
    \end{subfigure}
  \\
     \begin{subfigure}[b]{0.49\textwidth}
      \centering
       \includegraphics[width=1\textwidth]{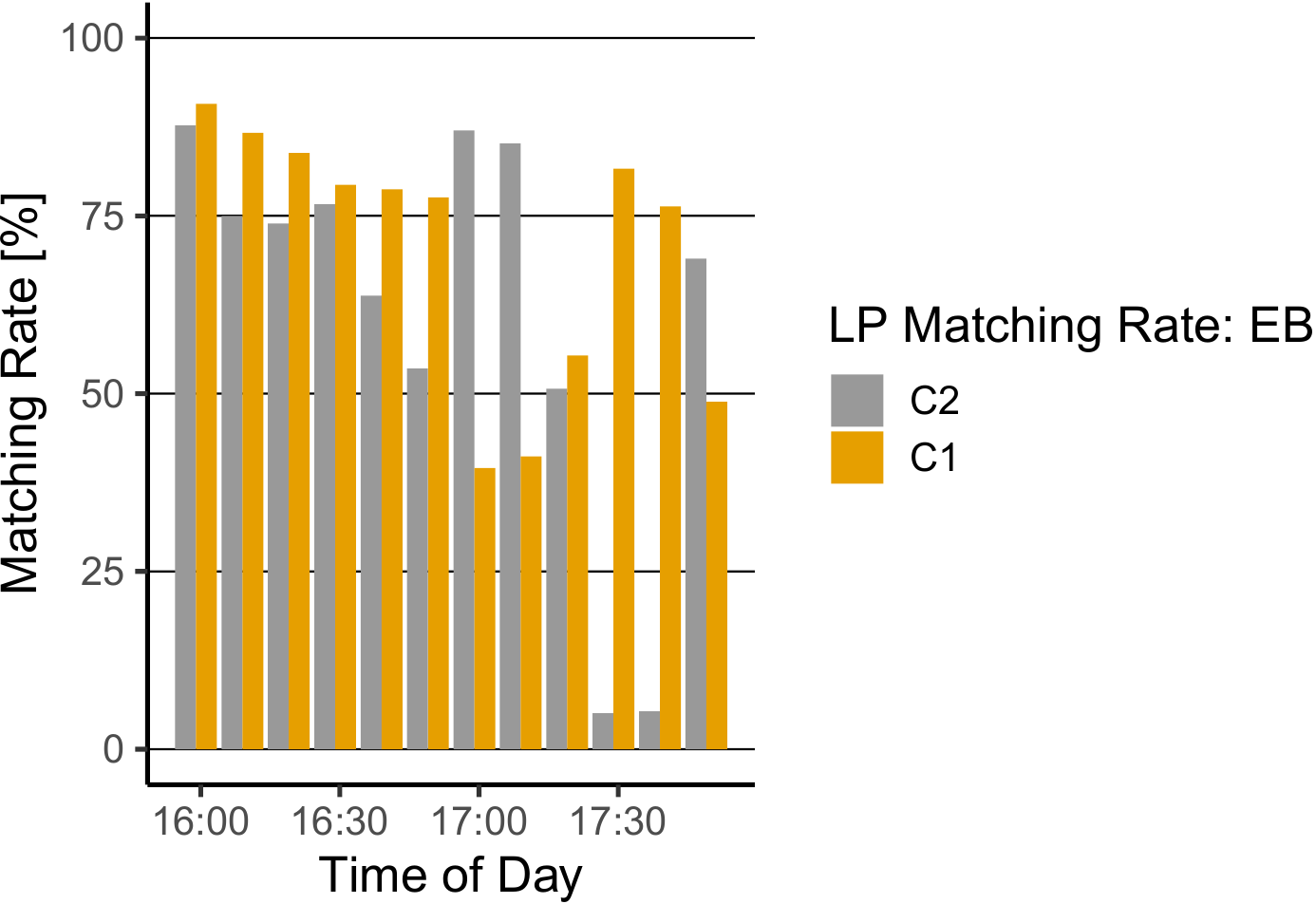}
        \caption{}
        \label{fig:matchq1q2}
    \end{subfigure}
    
    \caption{Flow evaluation of (a) $q_1(t)$ and (b) $q_2(t)$ at WB from the ground-truth measurement, the ALPR, and the thermal camera T2; (c) shows the matching rate of the ALPR algorithm for both derived flows.}
    \label{fig:flow_q1q2}
\end{figure}

\begin{figure}[t]
    \centering
      \begin{subfigure}[b]{0.49\textwidth}
      \centering
       \includegraphics[width=1\textwidth]{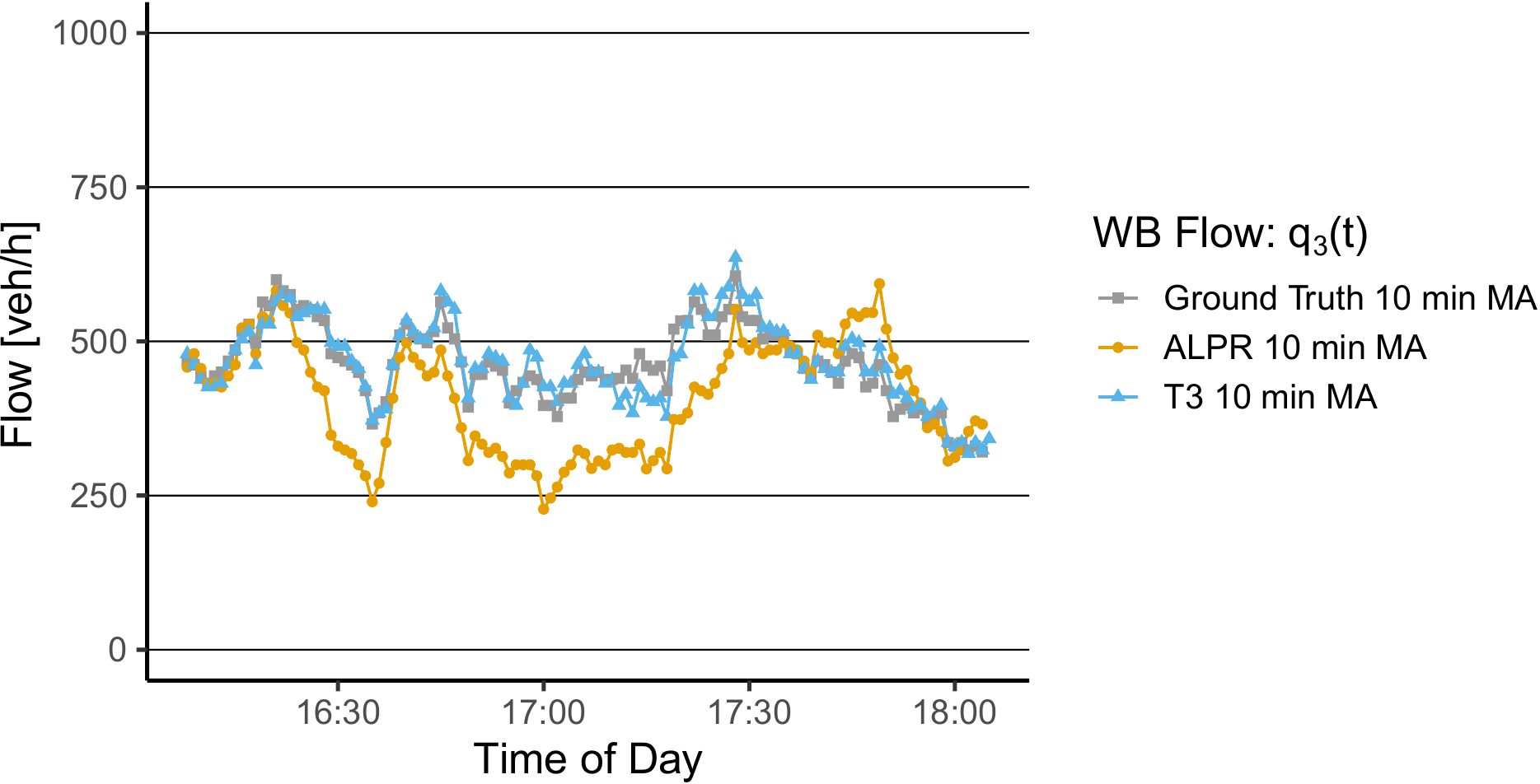}
        \caption{}
        \label{fig:q3}
    \end{subfigure}
    \begin{subfigure}[b]{0.49\textwidth}
    \centering
       \includegraphics[width=1\textwidth]{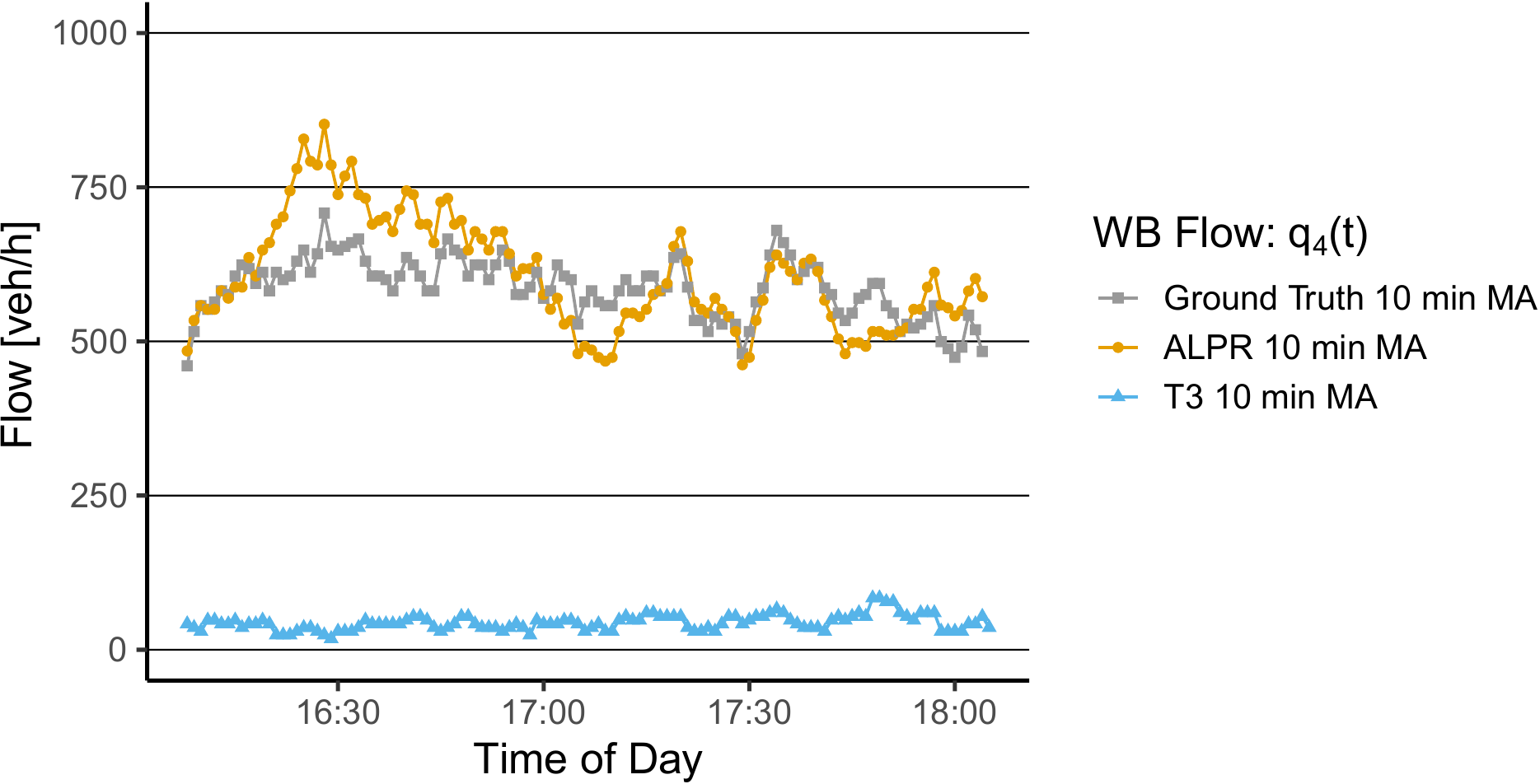}
        \caption{}
        \label{fig:q4}
    \end{subfigure}
   \\
    \begin{subfigure}[b]{0.49\textwidth}
    \centering
       \includegraphics[width=1\textwidth]{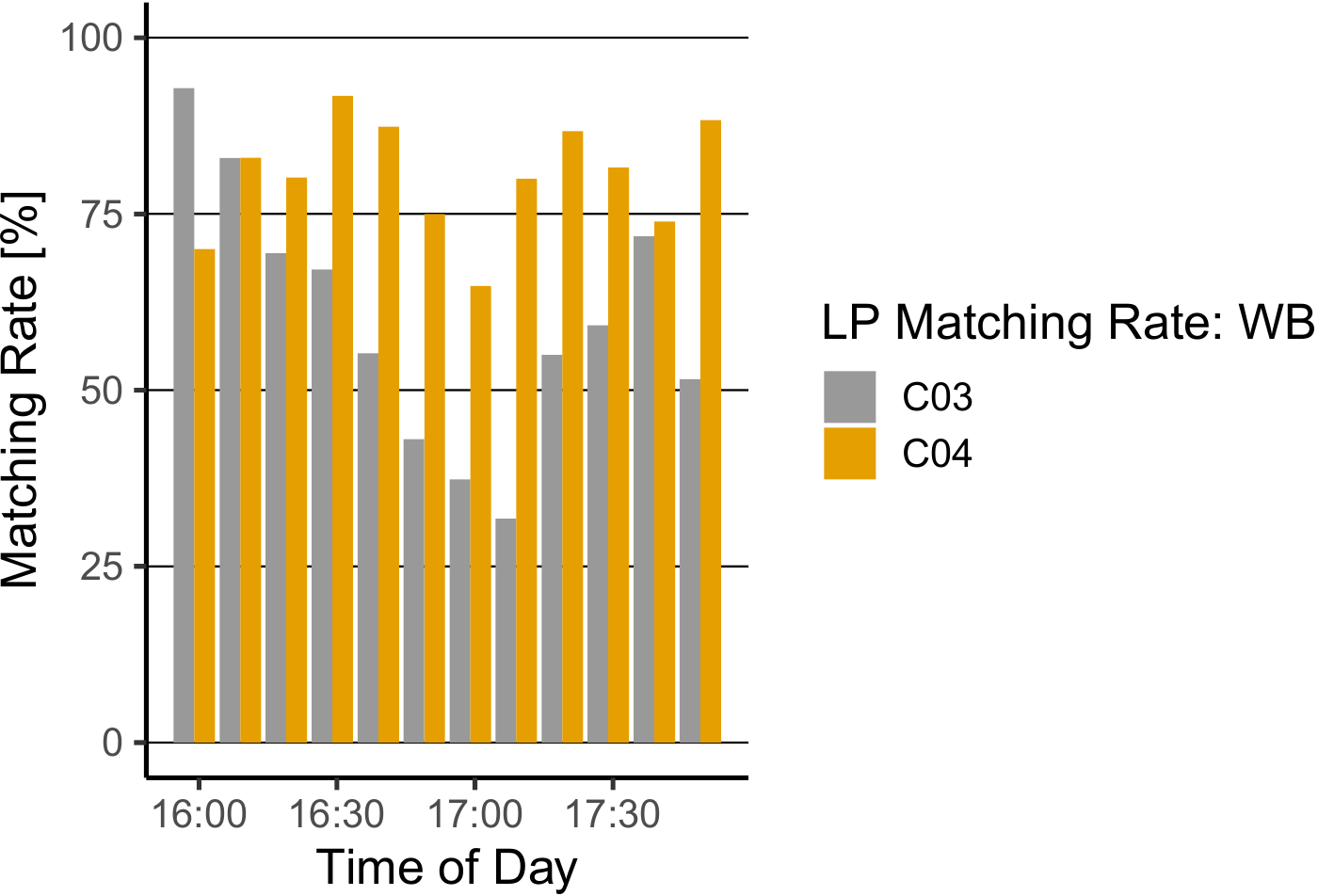}
        \caption{}
        \label{fig:matchq3q4}
    \end{subfigure}
    
    \caption{Flow evaluation (a) $q_3(t)$ and (b) $q_4(t)$ at EB from the ground-truth measurement, the ALPR, and the thermal camera T3; (c) shows the matching rate of the ALPR algorithm for both derived flows.}
    \label{fig:flow_q3q4}
\end{figure}

 The traffic flow results are derived by determining the vehicle counts from (a) the installed thermal cameras (b) the detected vehicles from the ALPR algorithm, and (c) the empirical measurement, i.e., the ground-truth data set. Note that the Google Distance Matrix does not allow the derivation of traffic flow. We derive all data sets and calculate the 10 minutes moving average of all time series $q_1(t)$ -- $q_6(t)$, i.e., the window size $k=10$. Figure~\ref{fig:q1} and Figure~\ref{fig:q2} present the derived flow time series $q_1(t)$ and $q_2(t)$ for all data sets. Also, Figure~\ref{fig:matchq1q2} denotes the matching rate of the ALPR algorithm. The matching rate is calculated by the fraction of detected vehicles by ALPR and the actually vehicle count from the ground-truth data set. 
 \label{sec:flow_results}
A comparison $q_1(t)$ and $q_2(t)$ of the ground-truth (in gray) and the data from the thermal cameras (in blue) T4 and T1 show a high correlation between the time series. Only small deviations of the thermal cameras' detection is noticeable. A quantitative analysis of the correlation coefficient $\rho$ and the $\mathrm{MAPE}$ show a coefficient result of 0.91 and an error rate of 4.83\% for $q_1(t)$ and $\rho=0.93$, $\mathrm{MAPE}=3.33\%$ for $q_2(t)$ (Table~\ref{tab:summary_flow_errors}). Further, the ALPR results are depicted in orange. The ALPR algorithm shows a good fit for $q_1(t)$ until the flow drops around timestamp 17:00 significantly. This can also be seen in the matching rate of C1 that drops below 50\%. After 17:30, the performance increases again with a matching rate around 70\% to 75\%. The average matching rate of C1 is 70\%, $\rho = 0.81$, and $\mathrm{MAPE} = 15.54\%$ (Table~\ref{tab:summary_flow_errors}). For $q_2(t)$ the ALPR shows a lower performance compared to $q_1(t)$. Several significant deviations from the ground-truth data can be observed throughout the investigated period. The matching rate of C2 shows an average rate of 59\%. Nevertheless, it is determined that the matching rate around 17:30 is below 10\%, which could be identified due to distortion caused by sunlight. The error metrics show a performance of $\rho = 0.61$, and $\mathrm{MAPE} = 12.64\%$. Figure~\ref{fig:flow_q3q4} depicts the results for all data sources of $q_3(t)$ and $q_4(t)$, respectively. 
 
\begin{figure}[!t]
    \centering
      \begin{subfigure}[b]{0.49\textwidth}
      \centering
       \includegraphics[width=1\textwidth]{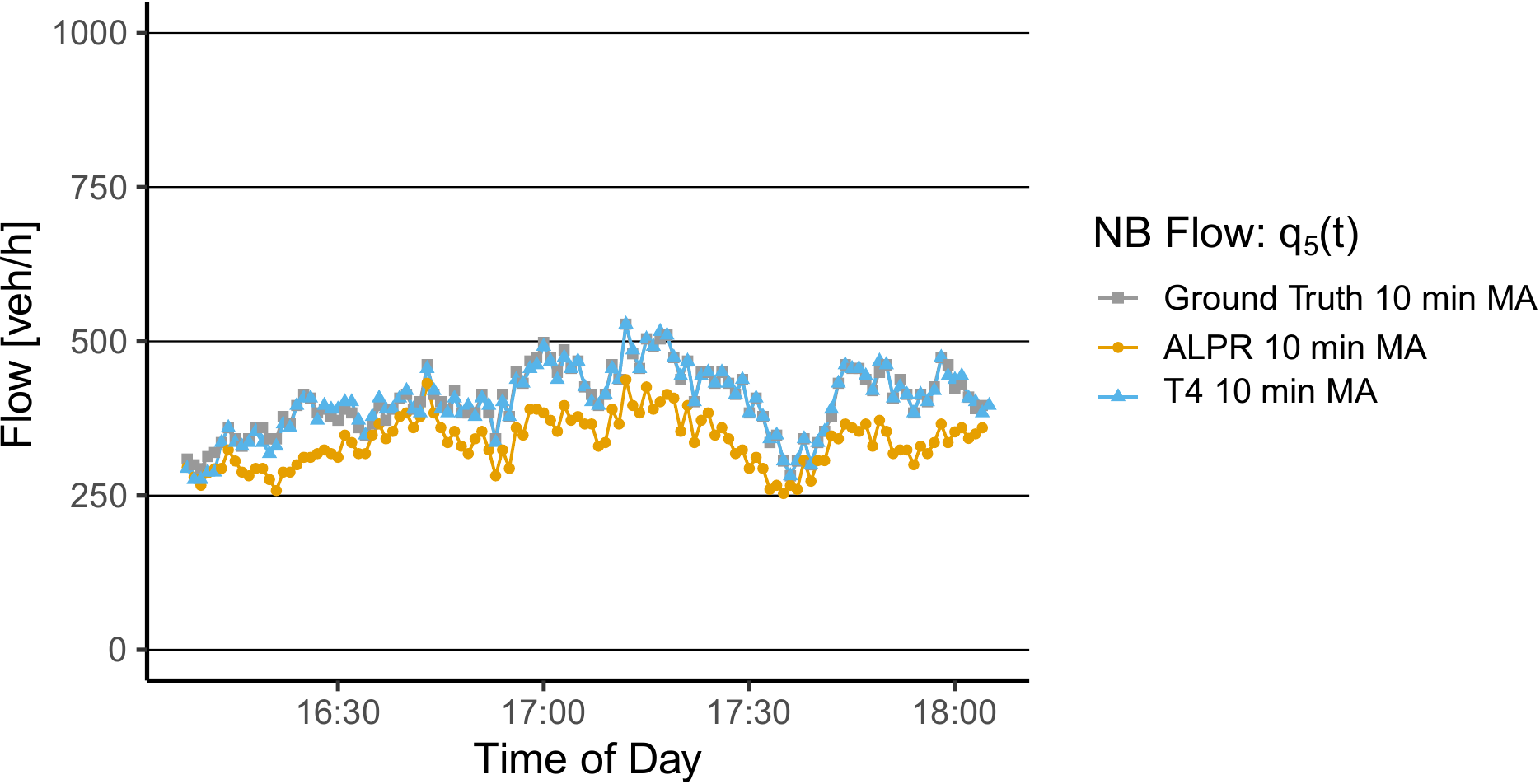}
        \caption{}
        \label{fig:q5}
    \end{subfigure}
    \begin{subfigure}[b]{0.49\textwidth}
    \centering
       \includegraphics[width=1\textwidth]{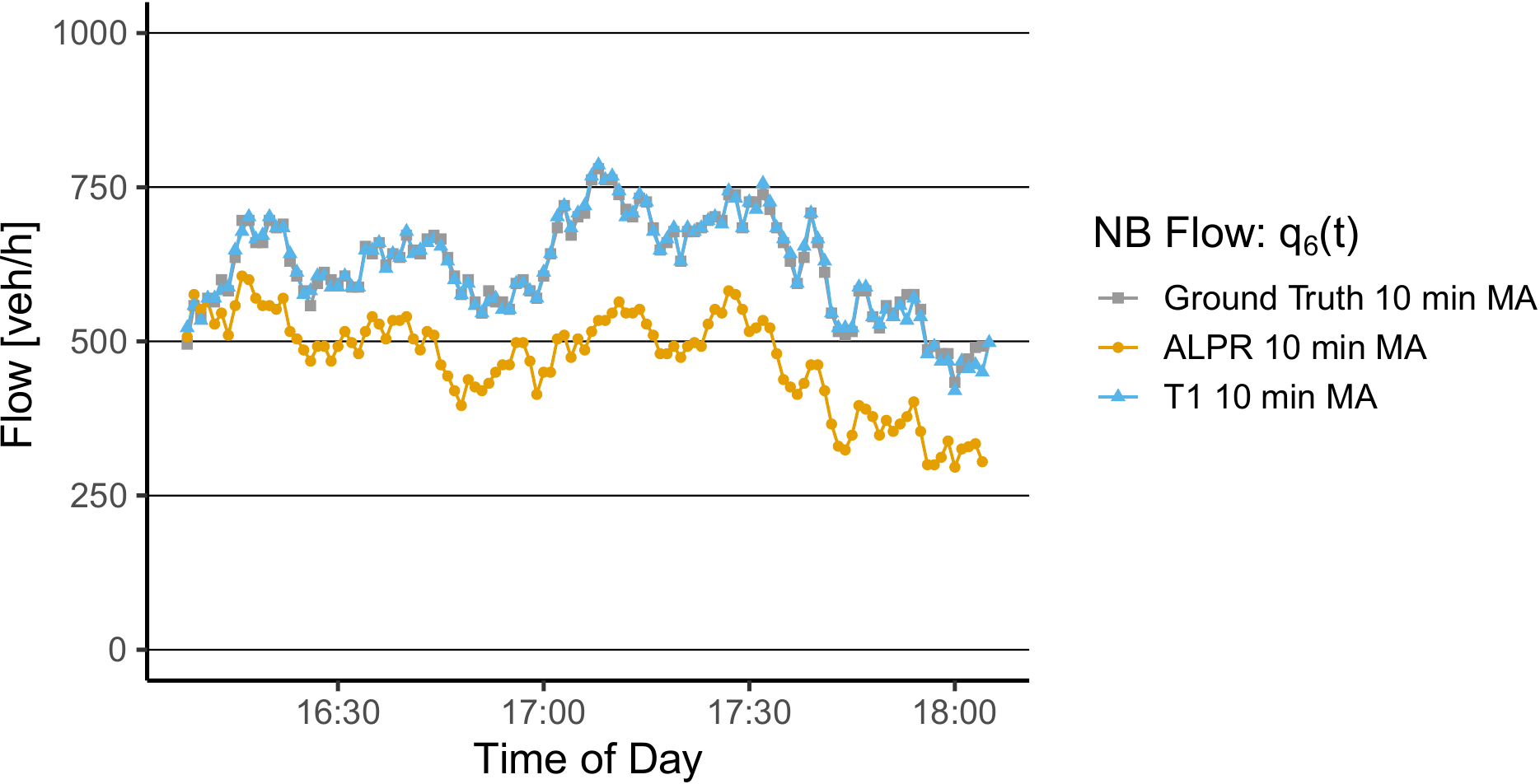}
        \caption{}
        \label{fig:q6}
    \end{subfigure}
    \\
     \begin{subfigure}[b]{0.49\textwidth}
    \centering
       \includegraphics[width=1\textwidth]{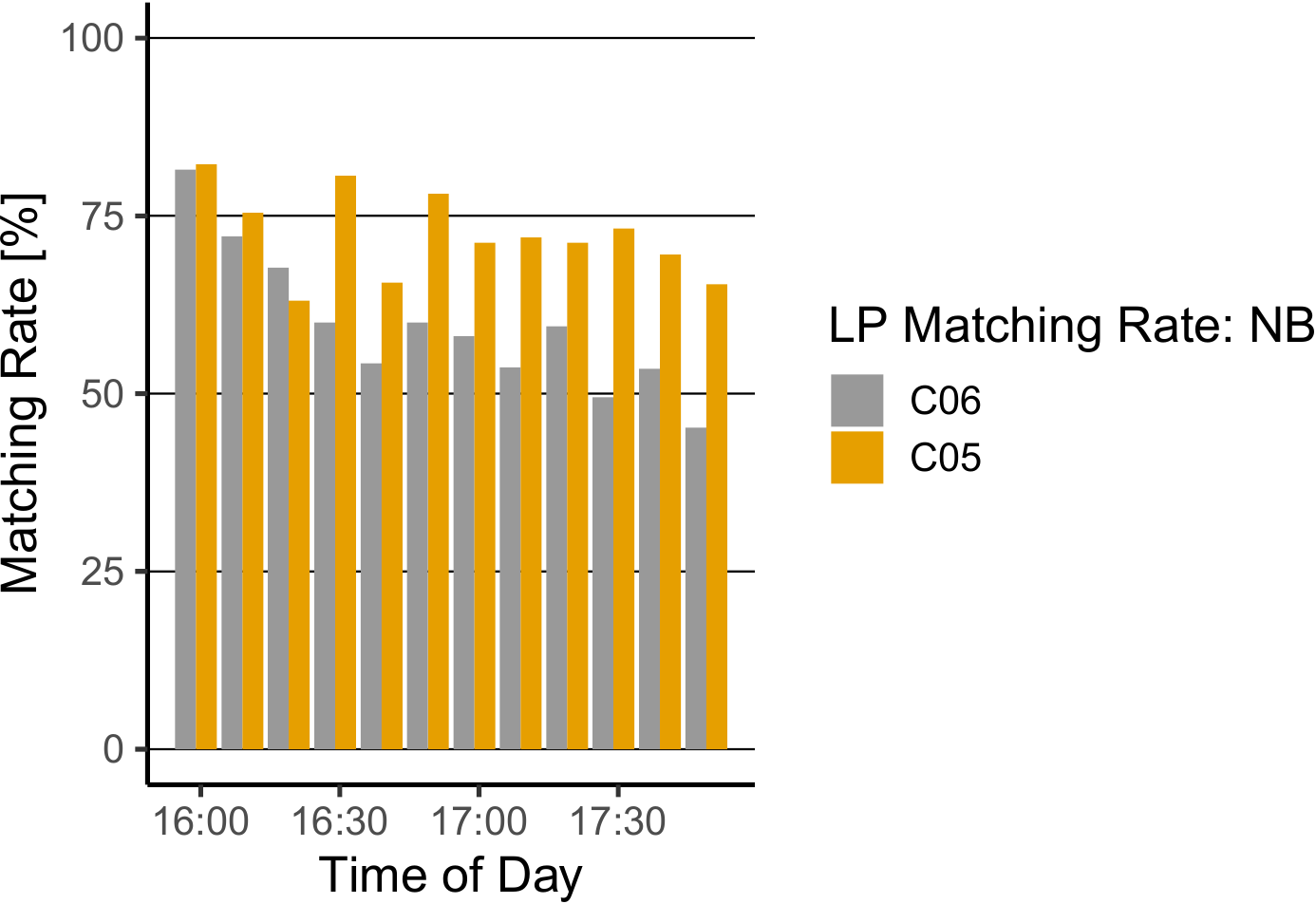}
        \caption{}
        \label{fig:matchq5q6}
    \end{subfigure}
    
    \caption{Flow evaluation (a) $q_5(t)$ and (b) $q_6(t)$ at NB from the ground-truth measurement, the ALPR, and the thermal camera T4 and T1; (c) shows the matching rate of the ALPR algorithm for both derived flows.}
    \label{fig:flow_q5q6}
\end{figure}

Again, a good fit of $q_3(t)$'s ground-truth data and the thermal camera data from T3 can be observed ($\rho = 0.94$, $\mathrm{MAPE}=3.86\%$). The ALPR results show significant deviations from the ground truth over time. Especially around 16:20, a drop in the flow is observed. Also, from 16:45 until 17:30, deviations are observed and are supported by the matching rate (Figure~\ref{fig:matchq3q4}) dropping below 50\%. The average matching rate is calculated with 59\%, $\rho=0.58$, and $\mathrm{MAPE}=17.20\%$. Figure~\ref{fig:q4} shows a high deviation from the ground-truth of the data set derived from T3. No correlation between the ground-truth and the thermal camera data is determined, i.e., $\rho=0.00$ and $\mathrm{MAPE}=92.45\%$. The reason for the high deviation is that T3 observes both traffic directions with one camera. Due to the high lane width (two lanes for individual transport and two additional lanes for public transportation), T3 cannot observe the traffic flow in this direction. A re-positioning of the camera or installing a second thermal camera could help improve the results. The ALPR shows a correlation of $\rho=0.73$ and $\mathrm{MAPE}=9.17\%$. The average matching rate is calculated with 76\%. The performance metrics are collected in Table~\ref{tab:summary_flow_errors}. Figure~\ref{fig:flow_q5q6} shows the results for $q_5(t)$ and $q_6(t)$.

\begin{table}[!t]
\caption{Correlation coefficient $\rho$ and MAPE of ALPR (abbreviation LP), thermal camera data (abbreviation TC) and the 10 min MA Ground Truth flow data.}
\label{tab:summary_flow_errors}%
\begin{tabular}{p{1cm}rrrrrrrrrrrr}
\toprule
              & \multicolumn{2}{c}{$q_1(t)$} & \multicolumn{2}{c}{$q_2(t)$} & \multicolumn{2}{c}{$q_3(t)$} & \multicolumn{2}{c}{$q_4(t)$} & \multicolumn{2}{c}{$q_5(t)$} & \multicolumn{2}{c}{$q_6(t)$} \\
              \midrule
              & LP       & TC        & LP       & TC        & LP       & TC        & LP      & TC         & LP       & TC        & LP       & TC        \\
              \midrule
$\rho$ {[}-{]}   & 0.81       & 0.91      & 0.61       & 0.93      & 0.58       & 0.94      & 0.73      & 0.00       & 0.58       & 0.94      & 0.76       & 0.99      \\
$\mathrm{MAPE}$ {[}\%{]} & 15.54      & 4.83      & 12.64      & 3.33      & 17.20      & 3.86      & 9.17      & 92.45      & 17.20      & 3.86      & 24.42      & 1.22     \\
\bottomrule
\end{tabular}
\end{table}

The thermal cameras of T4 and T1 show good results for $q_5(t)$ and $q_6(t)$. Note that two thermal cameras are installed due to a tram line between the two traffic lanes. For the flow $q_5(t)$ the correlation $\rho=0.94$ and $\mathrm{MAPE}=3.86\%$; for $q_6(t)$, $\rho=0.99$ and $\mathrm{MAPE}=1.22\%$. The ALPR algorithm shows deviations for both quantities. For $q_5(t)$ the average matching rate is 70\%, $\rho=0.58$ and $\mathrm{MAPE}=17.20\%$. The results for $q_6(t)$ show decreased ALPR algorithm performance over time. An inspection of the ground-truth video material showed that this is caused by increasing light reflections over time. The average detection rate is equal to 57\%, $\rho=0.76$ and $\mathrm{MAPE}=24.42\%$. Again, the quantitative results are collected in Table~\ref{tab:summary_flow_errors}.

The travel time results are derived by determining the timestamp when a vehicle enters and exits the system. As data sources  (a) the installed thermal cameras, (b) the detected license plates from the ALPR algorithm (c) tracked data from the Google Distance Matrix API, and (d) the empirical measurement, i.e., the ground-truth data set are used. We derive all travel time data sets and calculate the 10 minutes weighted moving average (the window size $k=10$) of the routes $r_1$, $r_3$, $r_4$, and $r_5$; consequently, $\tau_1(t)$, $\tau_3(t)$, $\tau_4(t)$, and $\tau_5(t)$. Note that for $r_2$ and $r_6$, the ground-truth data set showed low traffic volume (7 and 19 vehicles for the measurement period, respectively). Thus, data gaps occur, and a comparison of different data sources would not lead to a representative result. Therefore, these two routes are excluded from the analysis.  

\begin{figure}[!b]
    \centering
      \begin{subfigure}[b]{0.49\textwidth}
      \centering
       \includegraphics[width=1\textwidth]{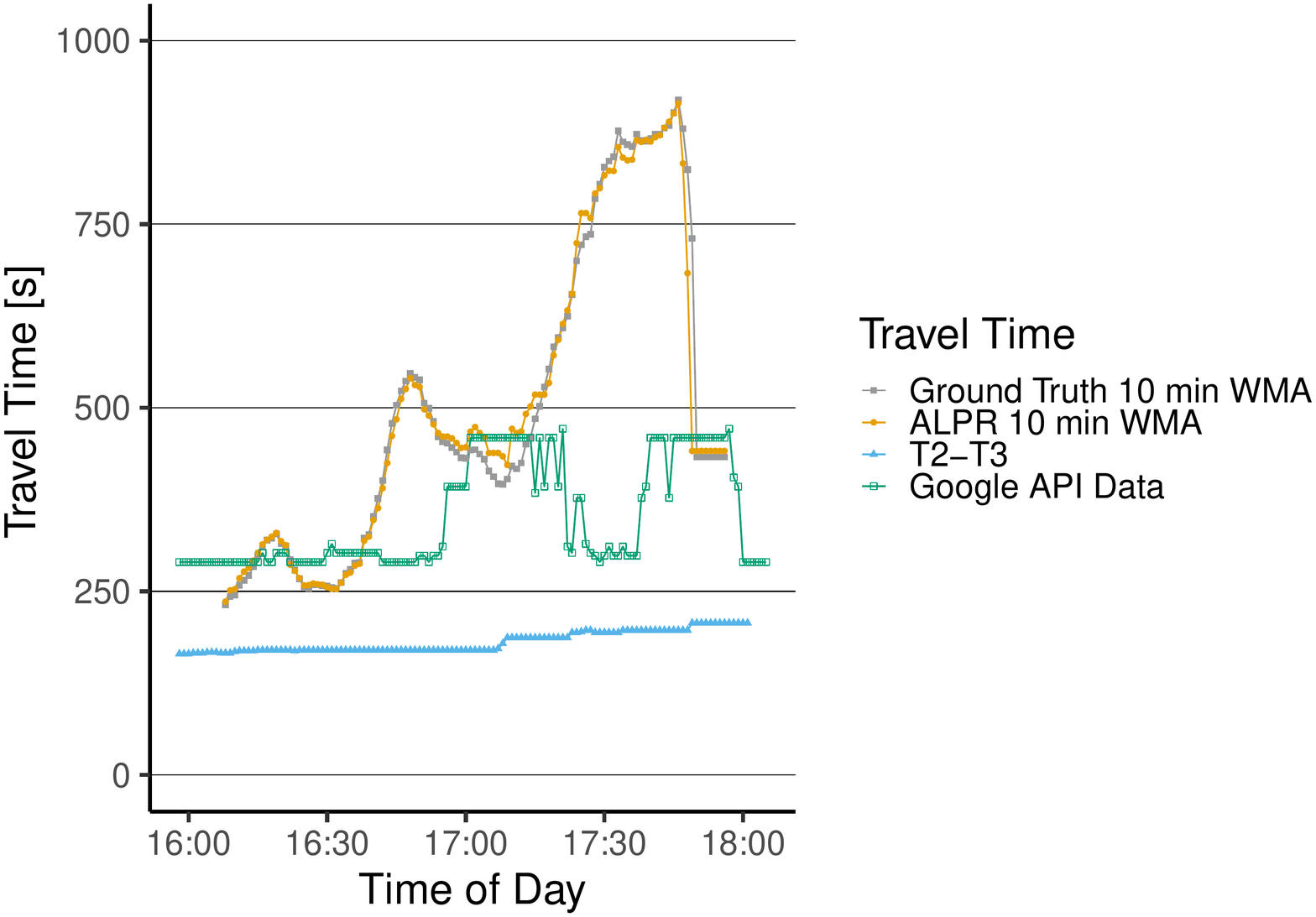}
        \caption{}
        \label{fig:tt_r1}
    \end{subfigure}
    \begin{subfigure}[b]{0.49\textwidth}
    \centering
       \includegraphics[width=1\textwidth]{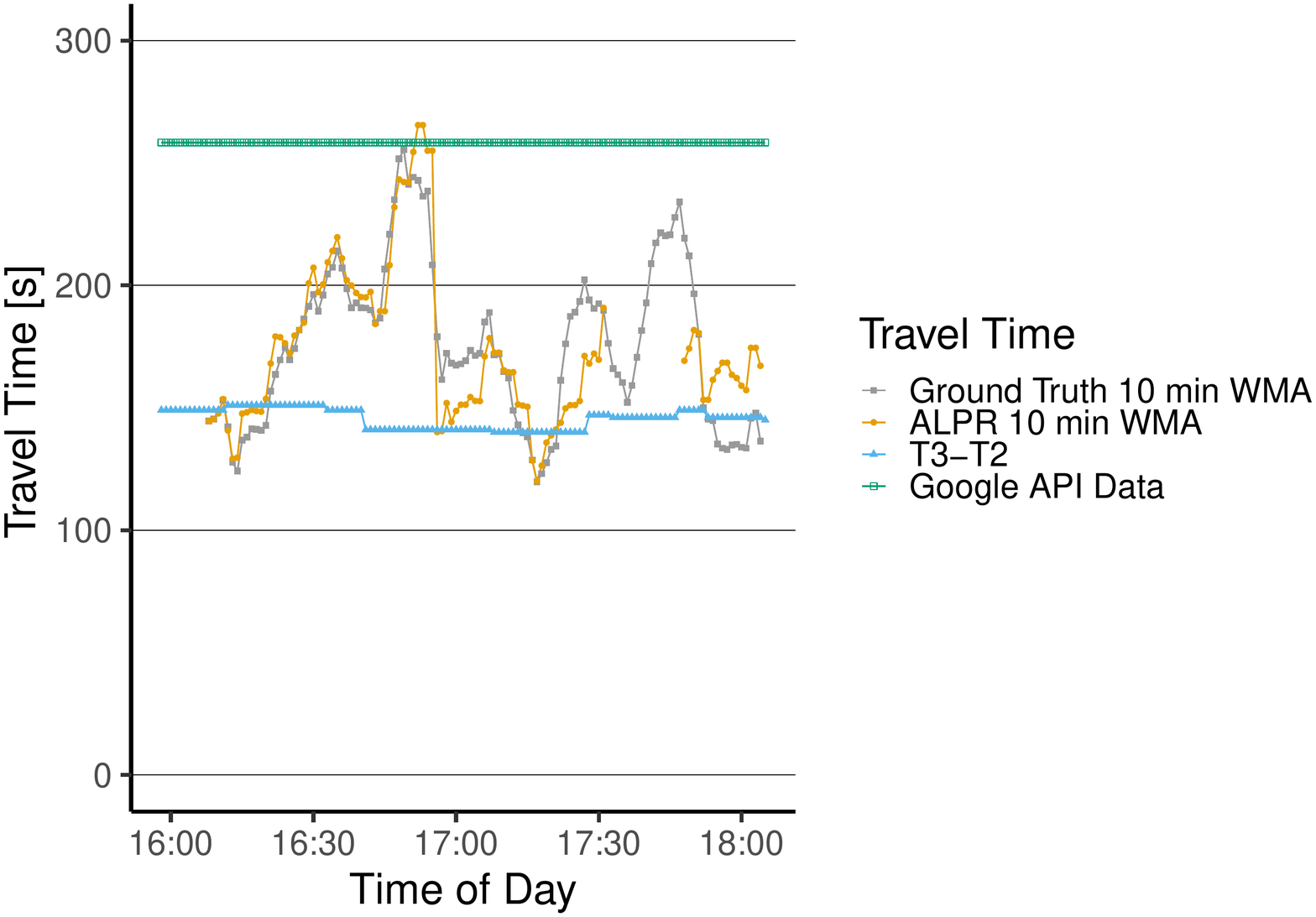}
        \caption{}
        \label{fig:tt_r3}
    \end{subfigure} \\
      \begin{subfigure}[b]{0.49\textwidth}
      \centering
       \includegraphics[width=1\textwidth]{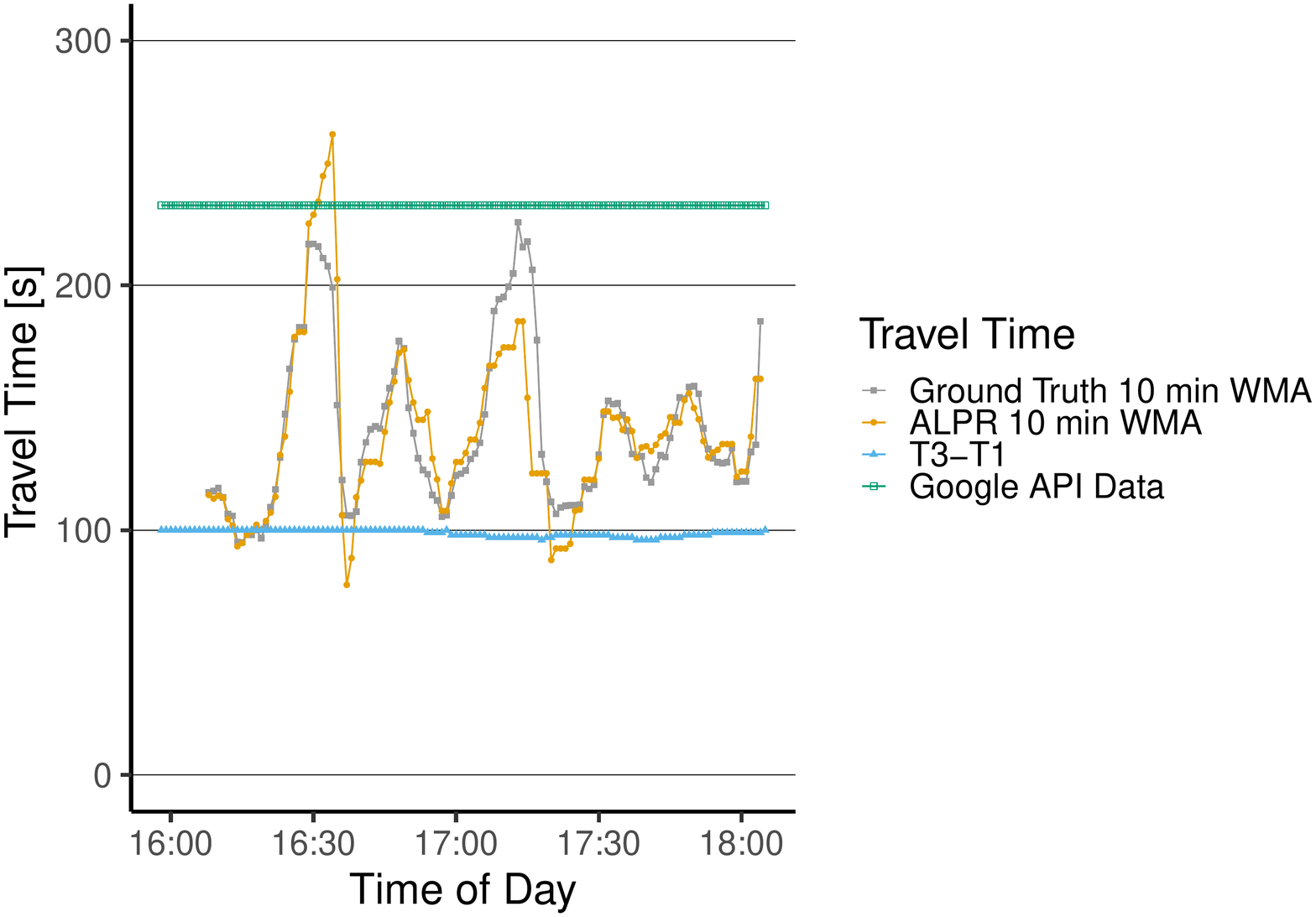}
        \caption{}
        \label{fig:tt_r4}
    \end{subfigure}
    \begin{subfigure}[b]{0.49\textwidth}
    \centering
       \includegraphics[width=1\textwidth]{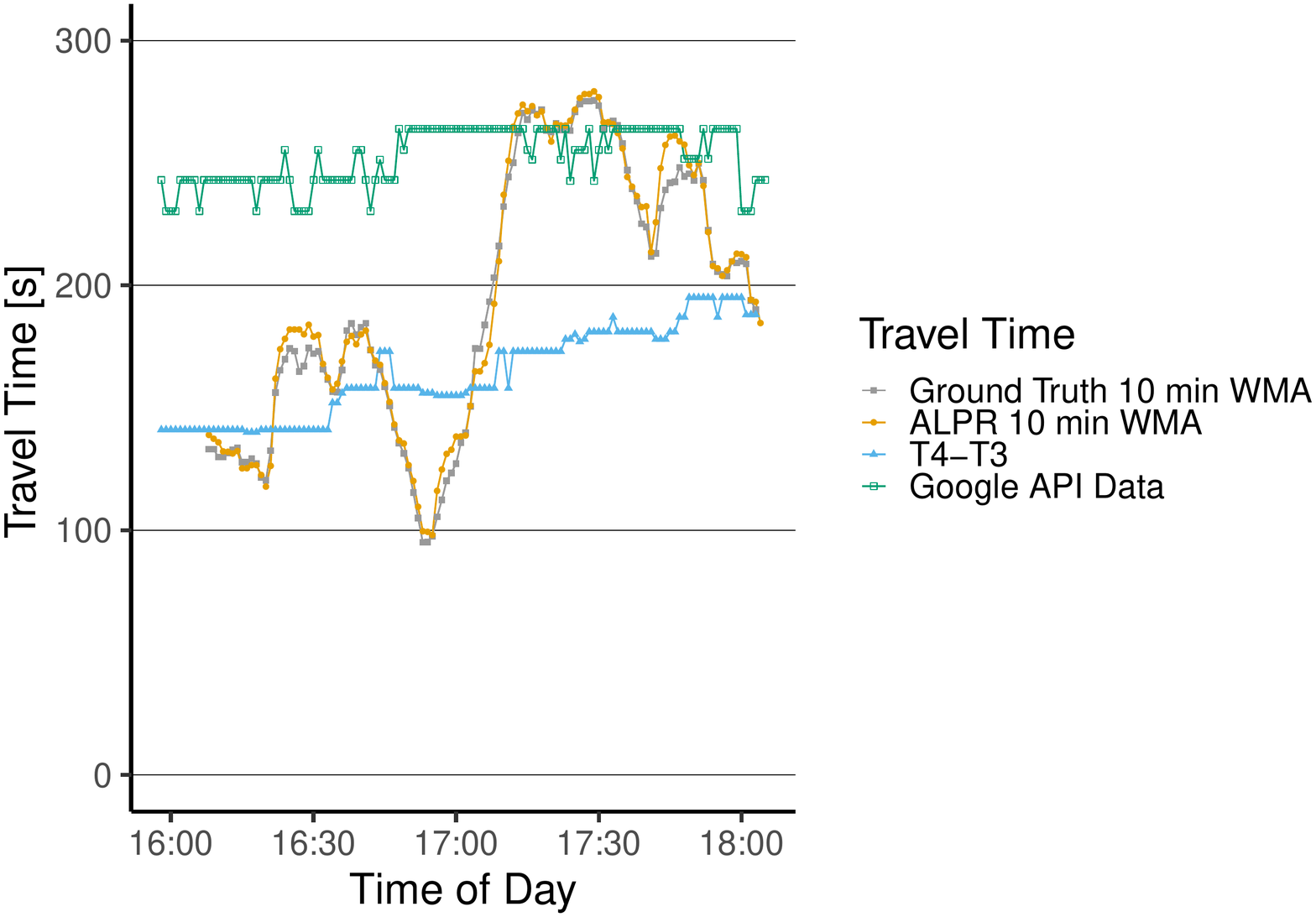}
        \caption{}
        \label{fig:tt_r5}
           \end{subfigure}
    \caption{Travel time evaluation for (a) $\tau_1(t)$, (b) $\tau_3(t)$, (c) $\tau_4(t)$ and (d) $\tau_5(t)$ with the ground-truth measurement, the ALPR, the set of thermal cameras, and the Google Distance Matrix API data set.}
    \label{fig:tt_r1_r3}
\end{figure}

Figures~\ref{fig:tt_r1} and~\ref{fig:tt_r3} present the derived travel time series $\tau_1(t)$ and $\tau_3(t)$ for all data sets. The travel time results for $\tau_4(t)$ and $\tau_5(t)$ are depicted in Figure~\ref{fig:tt_r4} and Figure~\ref{fig:tt_r5}.

The performance metrics are compiled in Table~\ref{tab:summary_TT_erros}. One can note that the quantity $\tau_1(t)$ increases over time, peaks around 17:45, and decreases again afterward. Results computed from the ALPR detections (orange time series) replicate this trend with small deviations. The time series correlate with $\rho=0.99$ and an $\mathrm{MAPE}=3.14\%$. Contrary results are shown by the set of thermal cameras T2 and T3 that are utilized for travel time derivation of $\tau_1(t)$. The time series only shows small variations and does not capture the trend of the ground-truth data ($\rho=0.71, \mathrm{MAPE}=58.07\%$). Potential reasons for the modest performance can be (a) a low penetration rate, i.e., a small number of WiFi devices are detected, or (b) the data is strongly post-processed. The time series computed from the Google Distance Matrix API shows a higher variance than the thermal camera data and fails to show the variation of the ground-truth data. Especially around 17:30, when travel time continues to rise, the Google Data (green time series) does not react to this system behavior. The correlation with the ground-truth results in $\rho=0.27$ and the $\mathrm{MAPE}=25.95\%$. The results show a similar trend for the travel time results of $\tau_3(t)$. The ALPR results replicate the ground-truth data with some small over- and underestimations. Nevertheless, one can note a data gap from 17:30 to 17:45, where no vehicles were detected. This results in a correlation of $\rho=0.84$ and $\mathrm{MAPE}=8.04\%$. Again the time series of the thermal camera data and the Google Distance Matrix API under- and overestimate the travel time, respectively. Both methods do not allow to react to a travel time change from, e.g., 180 sec to 250 sec at 16:45, as the peak is not covered. Additionally, the Google data does not show any variance, i.e., the standard deviation is zero. This also does not allow the determination of a correlation coefficient. Hence, Table~\ref{tab:summary_TT_erros} denotes such observations with 'NA'. For the thermal camera data, the computed performance metrics are $\rho=-0.11, \mathrm{MAPE}=18.36\%$, and for the Google data $\rho=\mathrm{NA}, \mathrm{MAPE}=50.64\%$.
For $\tau_4(t)$ similar results as for $\tau_3(t)$ are derived. The ALPR algorithm allows the derivation of travel times that show a good fit to the ground-truth ($\rho=0.85$, $\mathrm{MAPE}=7.51\%$) and the thermal camera data and also the Google data under- and overestimate the travel time with performance metrics of $\rho=-0.13, \mathrm{MAPE}=26.76\%$ and $\rho=\mathrm{NA}$, $\mathrm{MAPE}=73.38\%$, respectively. For the travel times on $r_6$, i.e., $\tau_6(t)$, the ALPR algorithm allows an accurate representation of the ground-truth data with $\rho=0.99$ and $\mathrm{MAPE}=2.73\%$. The thermal camera data shows a correlation of $\rho=0.72$ with an $\mathrm{MAPE}=20.60\%$ and the Google data allows the computation of $\rho=0.33$ and $\mathrm{MAPE}=45.49\%$.

\begin{table}[!t]
\caption{Correlation coefficient $\rho$ and MAPE of ALPR (abbreviation LP), thermal camera data (abbreviation TC), Google Distance Matrix API data (abbreviation G) and the 10 min MA Ground Truth flow data. Note that a correlation denoted as NA=not available means that the time series standard deviation is zero.}
\label{tab:summary_TT_erros}
\centering
\begin{tabular}{p{1cm}rrrrrrrrrrrr}
\toprule
 & \multicolumn{3}{c}{$\tau_1(t)$} & \multicolumn{3}{c}{$\tau_3(t)$} & \multicolumn{3}{c}{$\tau_4(t)$} & \multicolumn{3}{c}{$\tau_5(t)$} \\
 \midrule
                                                & LP   & TC     & G    & LP   & TC     & G    & LP   & TC     & G    & LP   & TC     & G    \\
                                                \midrule
$\rho$ {[}-{]}                  & 0.99   & 0.71   & 0.27  & 0.84   & -0.11  & NA     & 0.85   & -0.13  & NA     & 0.99   & 0.72   & 0.33   \\
$\mathrm{MAPE}$ {[}\%{]}        & 3.14   & 58.07  & 25.95    & 8.04   & 18.36  & 50.64  & 7.51   & 26.76  & 73.38  & 2.73   & 20.60  & 45.49 \\
\bottomrule
\end{tabular}
\end{table}

\subsection{Travel time estimation assessment}
\label{sec:estimation_results}
\begin{figure}[!t]
    \centering
    \includegraphics[width=0.5\textwidth]{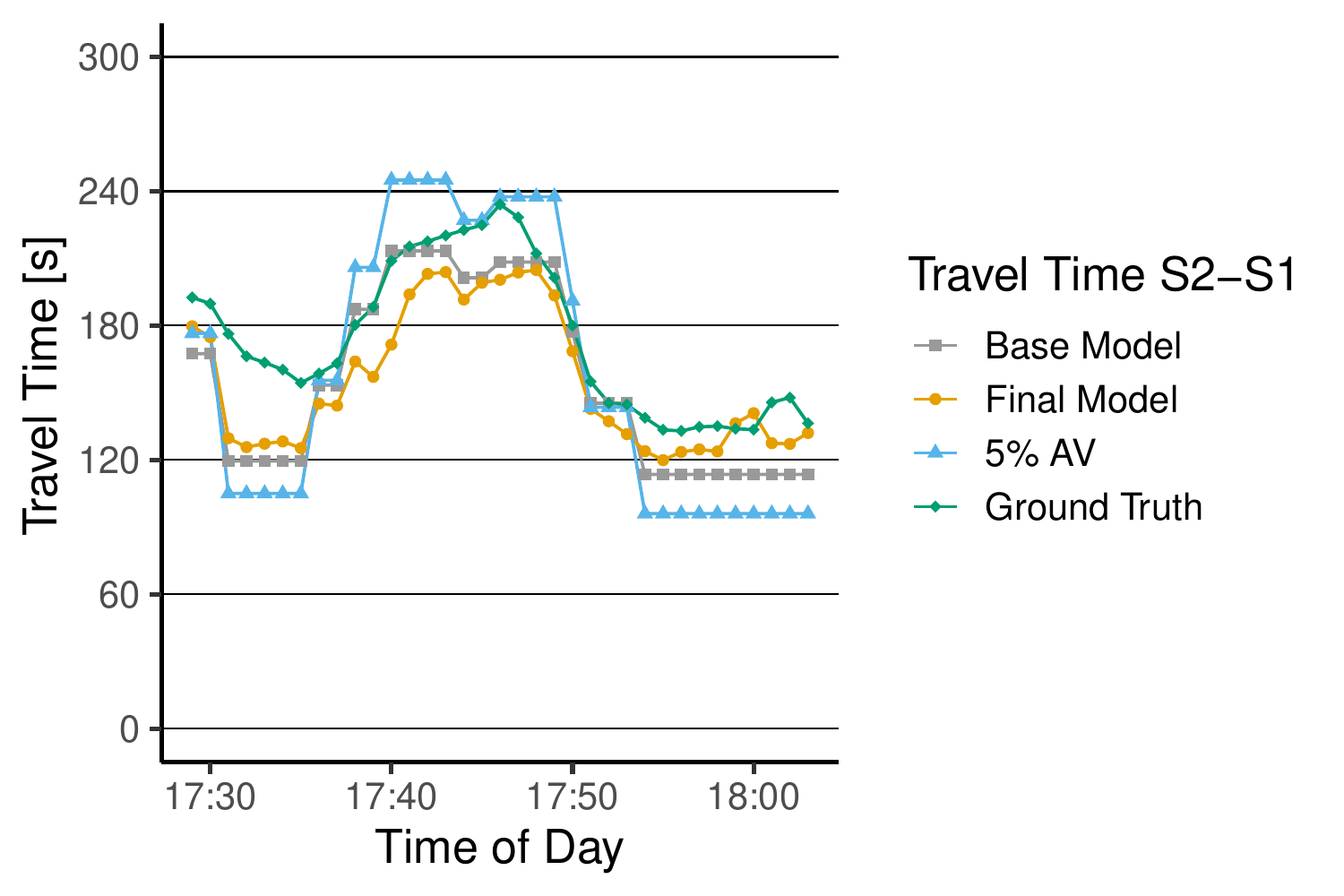}
    \caption{10 min WMA travel times estimates of the baseline model, estimation model,
compared to the 5\% data sample and the ground-truth on $r_3$.}
    \label{fig:estimation_result}
\end{figure}

To show the performance of a travel time estimation procedure, we train and apply the proposed MLR model. First, baseline models are trained with the data sample of (5\%) available moving sensor data as the only predictor for all routes $\mathcal{R}$. After creating the baseline, the other features are included in the model, and a final estimation model is derived. To show the performance of the trained model, we utilize $r_3$ as a test. Figure~\ref{fig:estimation_result} shows the comparison of the following time series: the actual ground-truth, the 5\% data sample without estimation, the travel time estimation of the baseline, and the final model. 

Results show that a 5\% data sample is insufficient to represent the ground-truth travel time $\tau_3(t)$. This is supported by a high $\mathrm{MAPE}=18.10\%$. The trained baseline model shows an $\mathrm{adj}R^2$ of 0.40 and over- and underestimates the travel time in Figure~\ref{fig:estimation_result}. The predicted time series results in an $\mathrm{MAPE}$ of 11.62\%. Nevertheless, it can be shown that our model already improves the prediction by 6.48\%. Finally, we apply the model with all features utilized as predictors. Although the model also indicates deviations from the ground-truth data, the $\mathrm{adj}R^2$ equals 0.81, and the $\mathrm{MAPE}$ reduces further to 10.92\% (see Table~\ref{tab:predicton_performance_r3}). 

\begin{table}[t]
\caption{Adjusted R-square values of the model performance on the training data. MAPE denotes the comparison of travel times estimates (baseline model and estimation model), and 5\% sample of $r_3$.}
\label{tab:predicton_performance_r3}%
\centering
\begin{tabular}{lccc}
\toprule
$r_3$    & $\mathrm{adj}R^2$ {[}-{]} & MAPE {[}\%{]}  \\ \midrule
5\% sample  & -                           & 18.10                                \\
Base Model     & 0.40                       & 11.62                           \\ 
Final Model    & 0.81                        & 10.92                          \\
 \bottomrule
\end{tabular}

\end{table}

\section{Conclusion}
\label{sec:conclusion}
The paper presents insights on arterial roads traffic state representation in terms of traffic flows and travel times in an urban transportation network. Additionally, a methodology for travel time estimation is proposed. For assessment purposes an experimental campaign with video measurements within a restricted area in Zurich, Switzerland has been organized, while additional data sources (thermal cameras, loop detector data, signaling data) were made available to the authors by the city of Zurich.

More specifically, the work investigates a sensor-based assessment and takes into account (a) data from thermal cameras, (b) post-processed video data with an ALPR and (c) travel times from the Google Distance Matrix API. All post-processed data sources are then compared to an empirical ground-truth measurement that is conducted in a particular area in Zurich, Switzerland. Performed traffic measurements with video cameras allow compiling a data set with true traffic flow of six detection spots and travel time values of six routes during two hours (afternoon peak). Results show that for the derivation of traffic flow, the thermal camera performs best with an $\mathrm{MAPE}$ below 5\% for all detection spots. The ALPR algorithm shows error values between 9\% and 25\%, which are due to decreasing detection rates of the procedure caused by, e.g., light reflections. For the derivation of travel times, the ALPR outperforms all other sensor technologies with an $\mathrm{MAPE}$ below 8\% for all routes. However, the performance of data from thermal cameras and the Google Distance Matrix API is modest, with error values between 18\% and 58\%. Finally, we showcase the performance of an MLR model for travel time estimation. The model architecture gets several extracted features from loop detector data and a 5\% data sample from moving sensor data (e.g., CAVs) as an input. We compare the model performance with (a) the 5\% data sample alone, (b) a baseline model that only sees the 5\% data sample for training, and (c) the actual ground-truth data. Results show that a data sample such as 5\% of the ground-truth data is not enough to represent the travel time for a specific route. With the MLR baseline model, the $\mathrm{MAPE}$ can be reduced from 18.10\% to 11.62\%; with the final model, the error reduces further to 10.92\%. 
Future research will focus on more detailed modeling approaches for travel time estimation in the area, e.g., Kalman filtering. Additionally, it should be investigated how thermal cameras performance can be improved so that the sensors provide good quality for traffic flow and travel time derivation.


\bibliographystyle{trb}
\setlength{\bibsep}{0pt}
\bibliography{bib}
\section*{Author Contribution Statement}
The authors confirm contribution to the paper as follows: study conception and design:
 Noel Hautle, Alexander Genser, Michail Makridis, Anastasios Kouvelas; analysis and interpretation of results: Noel Hautle, Alexander Genser; draft manuscript preparation: Alexander Genser, Michail Makridis, Anastasios Kouvelas. All authors reviewed the results and approved the final version of the manuscript.
\end{document}